\documentclass[]{aa}  
\usepackage{graphicx}
\usepackage{txfonts}
\usepackage{natbib}
\usepackage{threeparttable}
\usepackage{epsfig}

\newcommand{\stf}{{\textsl{StarFinder }}}
\begin{document}
   \title{Searching for substellar companions of young isolated neutron stars\thanks{Based on observations made with ESO Telescopes at the La Silla or Paranal Observatories under programme IDs: 66.D-0135, 71.C-0189, 72.C-0051, 74.C-0596, 077.C-0162, 78.C-0686, 79.C-0570 }}

   \author{B.Posselt
          \inst{1,2}
          \and
          R. Neuh\"{a}user
	  \inst{1}
	  \and
	  F. Haberl
	  \inst{2}
          }

   \offprints{B.Posselt}

   \institute{Astrophysikalisches Institut und Universit\"{a}ts-Sternwarte, Schillerg\"{a}\ss chen 2-3, 07745 Jena, Germany\\
         now at the CfA -- \email{bposselt@cfa.harvard.edu}
         \and
             Max-Planck-Institut f\"{u}r extraterrestrische Physik, Postfach 1312
85741 Garching, Germany \\ 
             }

   \date{}

 
  \abstract
   {Only two planetary systems around old ms-pulsars are currently known. 
   Young radio pulsars and radio-quiet neutron stars cannot be covered by the usually-applied radio pulse timing technique. However, finding substellar companions around these neutron stars would be of great interest -- not only because of the companion's possible exotic formation but also due to the potential access to neutron star physics.}
   {We investigate the closest young neutron stars to search for substellar companions around them.}
   {Young, thus warm substellar companions are visible in the Near Infrared while the neutron star itself is much fainter. Four young neutron stars are moving fast enough to enable a common proper motion search for substellar companions within few years. }
   {For Geminga, RX~J0720.4-3125, RX~J1856.6-3754, and PSR~J1932+1059 we did not find any co-moving companion down to 12, 15, 11, 42 Jupiter masses for assumed ages of 1, 1, 1, 3.1~Myrs and distances of 250, 361, 167, 361\,pc, respectively. 
   Near Infrared limits are presented for these four as well as five other neutron stars for which we    currently have only observations at one epoch.}
   {We conclude that young isolated neutron stars rarely have brown
dwarf companions.}


   \maketitle
%

\section{Introduction}

For a long time substellar companions (planets and brown dwarfs) around neutron stars have not been expected due to the powerful supernova giving birth to the compact object.
Astonishingly however, the first planetary mass companions ever discovered
around other stars were neutron star (NS) companions. Three such objects
  are confirmed around the old millisecond(ms)-pulsar PSR~1257+12
\citep{WF92,Konacki2003}. They have up to four~Earth masses and are very close
to the primary -- 0.2~AU to around 0.5~AU. 
Another planetary system around a NS is  the binary system
of the ms-pulsar PSR~B1620-26 and its white dwarf companion \citep{Sig03,Sig05}. A 2.5~Jupiter mass object is orbiting the white dwarf at 23~AU and thus also the NS.
Currently, we do not know of any brown dwarf companion around a neutron star.
Several formation models have been discussed to explain the unexpected discovery of NS planets either by assuming supernova survival, by planet formation after the supernova or by planet capturing (for a detailed review see \citealt{Pod95}).
The planet formation from in-situ disks around NSs could differ from those around solar-like stars.  Thus, the study of NS planetary systems in comparison
to those around solar-like stars can be very fruitful for the
understanding of planet formation in general.
But substellar companions  of NSs are not only interesting for planet or brown dwarf formation theory. They could also provide insights into the supernova event and maybe the composition and final stages of the progenitor star(s). 
Studying low-mass companions around NSs, especially the companion masses and  their orbital characteristics, can determine also the dynamical masses of the NSs.
These masses, together with the star radii, constrain the equation of state of matter in the neutron stars. Currently, no radius and mass of a neutron star are known precisely for the same object and the equation
of state of ultra-dense matter in the universe is still not known. Whereas masses of NSs can be derived for binary systems (e.g. \citealt{Jonker03}), the best radius estimation comes from the radio-quiet X-ray thermal NSs, which are isolated without stellar companions (e.g. \citealt{Trumper04}).\\
Only the two planetary systems around ms-pulsars mentioned above 
were found so far, both by the radio pulse
timing technique, which measures variations in the pulse arrival times.
This technique is most sensitive to very stable ms-pulsars and requires much observing time. For ordinary radio pulsars the timing noise hinders the precise measurements needed. 
Radio-quiet neutron stars can not be searched for substellar companions by this method.
The radio pulse timing technique can detect only relatively close companions like in radial velocity planet searches among normal solar-like stars. Therefore, substellar companions in wide orbits cannot be detected with this method.  
As NSs are very faint at visible wavelengths, the commonly used radial velocity method or the search for
astrometric wobbling should preferably be done in X-rays where many NSs are bright. 
Unfortunately the spectral and spatial resolution in X-ray astronomy
is currently not good enough for these methods. 
Transit searches would
not be effective, because rarely more than one
NS would be in one field of view.\\
Substellar companions of NSs at any separations can be found only
by their common proper motions detected by direct imaging, primarily in the near infrared (NIR) if the companions are young. 
They are several magnitudes fainter in the optical.
The companions are brighter than the NSs in the NIR, even though
lower in mass, because they are larger in size and still contracting and, hence, warm when young.
Direct detection of very close (even few AU) and faint substellar companions is
currently possible only around NSs. Substellar companions in wide orbits can be detected too.
However, one is restricted to young objects at small distances from the sun. 
The magnitudes of substellar objects (brown dwarfs and giant planets) can be predicted following the non-gray theory of substellar objects by \cite{Burrows1997} (B97 in the following), who have computed IR magnitudes of substellar objects depending on age and mass. 
The upper mass limit of planets is not yet defined.
There are basically two suggestions:
Either all substellar objects below the D burning limit at
around 13 Jupiter masses are referred to as planets or all objects
below the radial velocity brown dwarf desert (at
20 to 50 Jupiter masses) are called planets, i.e., either all
objects below 13 Jupiter masses or below 50 Jupiter masses.
Calculation like those by B97 start
with assumed initial conditions about the internal
structure and temperature of the substellar objects
and should, hence, be taken with care for very young
objects up to a few Myrs (for a discussion see \citealt{Baraffe2002}). 
However, we use these calculations for our estimates due to the lack of better suited ones.
According to B97's calculations, at an age of 100~Myrs and a distance of 300~pc, a companion at the Deuterium burning mass limit ($\sim 13$~M$_{\rm Jup}$)
would have an H-band magnitude of 22 mag -- easily achieved by current optical telescopes.
We present in the following the first results of our search for substellar companions around NSs. 
We note that part of our data have been analysed by
  \citet{Curto2007}. However, they were interested in the
  spectra of the neutron stars as well as in hypothetical fallback disks
  around them. They neither found a fallback disk nor was the IR data deep
  enough to constrain the spectra of the NSs.


\section{Sample selection}
\label{sample}
While substellar companions may survive the supernova (SN), this case is probably
rare, thus formation after the SN likely for most substellar companions. 
Hence, we use the NS age as good approximation of the age of possible
substellar companions. In case of SN survival the companion's age can be significantly underestimated.
All NSs younger than 100 Myrs and closer than 300~pc were chosen as the initial sample in 2000. 
The radio pulsars amongst these neutron stars were selected from the pulsar catalog of \citet{Taylor1993}. They were thought to be ordinary radio pulsars, thus still young as radio emission from non-recycled NSs is expected to last only at most 10~Myrs \citep{Handbook2004}.
To that sample, the radio-quiet X-ray thermal isolated neutron stars (XTINSs) were added. 
As noted above, the XTINSs are especially interesting to constrain the equation of state due to their blackbody-like spectra resulting in the best radius estimation currently available. They also represent nearly half of the local young neutron star population (for reviews on these NSs, see, e.g., \citealt{Kaplan2008,Haberl2004b}).
No close co-moving companions down to substellar masses have been identified
in deep optical observations and proper motion studies (e.g., \citealt{Mignani2003, Walter1997,Motch2003}) 

The ages of the radio pulsars in our sample are estimated roughly from the characteristic spin-down age. However, the characteristic age is only a very coarse estimation of the true neutron star age. For the well known Crab pulsar the spin-down age exceeds the true age by about $25\%$ (e.g., \citealt{Manchester77}). 
Initially assumed to be younger than $100$ Myrs, some of our selected radio
pulsars seem to be older from newer radio timing data by \citet{ATNF2004}, so that possible substellar companions would be fainter (see individual discussion in Section~\ref{result}).
The additional radio-quiet XTINSs found as soft
ROSAT sources are most probably also young NSs ($<100$\,Myrs): Their proper motions indicate
likely origins in nearby young star-forming clouds.
The XTINSs have high magnetic fields, an indicator of youth, and inferred spin-down ages from X-ray pulsations are less than 10~Myr (see, e.g., \citealt{Kaplan2008}). 
Thus, the XTINSs are probably several Myrs old, with cooling surfaces which emit the
X-ray emission. 
For all seven radio-quiet X-ray thermal NSs, it appears unlikely that they are old with their X-ray emission due to accretion from the interstellar medium (ISM). First, they have constant X-ray emission over years with one notable exception where precession of the NS is likely \citep{Haberl2006aph}. \cite{NT99} quoted also statistical reasons for excluding accretion from the ISM. 
Furthermore, three radio-quiet X-ray thermal NSs have known proper motions, 
which are so high that they imply too low Bondi-Hoyle accretion efficiencies 
to describe these stars as old ($10^9$) NSs re-heated by accretion from the ISM. 

For the radio pulsars, rough distance estimates are available from the dispersion
measure, which indicates the electron column density towards the pulsar  \citep{Taylor1993, Manchester2005}. 
For the radio-quiet X-ray thermal NSs, a parallax measurement is available only in two cases, namely RX~J0720.4-3125 and RX~J1856.5-3754. For the latter, \citet{Walter2002} determined a distance of $\sim 117$~pc with the {\textsl{Hubble Space Telescope}}. 
Recently \citet{Kaplan2007} reported an improved distance estimation with RX~J1856.5-3754 being at $167^{+18}_{-15}$~pc. 
For RX~J0720.4-3125 \citet{Kaplan2007} reported a parallactic distance of $361^{+172}_{-88}$~pc. 
The measured column densities from the X-ray spectra 
of the other XTINSs are similar to those of RX~J0720.4-3125 and RX~J1856.5-3754, all 
 smaller than the Galactic values (towards the respective direction of sight).
One can assume the same order of magnitude for their distances as for RX~J1856.5-3754 and RX~J0720.4-3125.
Taking into account absorption by the inhomogenously distributed ISM, the estimated distances for two further radio-quiet X-ray thermal NSs support this assumption \citep{Posselt2007}.
Here, we take 300~pc as $lower$ distance limit for the other 5 XTINSs.\\
In the following we will present the observations and results for nine neutron stars, four of them observed at two epochs.
All objects, their known or assumed distances, and their proper motions are listed in Tab.~\ref{objects}.\\
Given the age and distance $d$ of the NSs, the expected $H$-band magnitude to find brown dwarfs and massive gas planets can be derived applying the results by B97.
Bolometric corrections in the $K$-band have a reported range from $\approx 2$
(T5 dwarf) to  $\approx 3.5$  (L5 dwarf) \citep{Golimowski2004}.
We use in the following always a bolometric correction, $BC = BC_H \simeq BC_K =
3.5$\,mag.

\section{Observation summary and data reduction}
\label{obs}
\subsection{First epoch observations}
 Nine neutron stars were observed in $H$-band
 utilising the Very Large Telescope (VLT) equipped with {\it{Infrared
    Spectrometer And Array Camera}} ($ISAAC$) in service mode.
The objects and the details of observations are listed in Tables~\ref{objects} and
\ref{obs}  in Sec.~\ref{detailsobs} of the  Online Appendix.

The VLT observation durations ranged from two to three hours. The seeing was usually better than $0.8 \arcsec$ and the airmass in most cases below 1.5. The $jitter$-Mode was used with individual exposures of 12~sec and a jitter box width of $20 \arcsec$.
The data have been reduced with ESO's $Eclipse$ (version 4.8.1 and 5.0.) \citep{eclipse1999,eclipswe2001} taking into account the corresponding darks, sky flats and bad pixel maps obtained during ESO's regular calibration plan. 
\begin{threeparttable}[t]
\caption[The observed objects]{\textbf{The observed neutron stars} \protect\\  Listed are for each object: the distance $D$, the proper motion in Right ascension ($\mu_{\alpha}$) and Declination ($\mu_{\delta}$) if known ($n.k.$ stands for $not$ $known$). 
}
\label{objects}
\begin {tabular}{@{} l    c  c  c @{}}
\hline
Object  & $D$ & $\mu_{\alpha}$ & $\mu_{\delta}$ \\
   &  [pc] & [mas yr$^{-1}$]& [mas yr$^{-1}$] \\
   &    &                &                 \\
\hline

PSR J0108-1431 &   130\tnote{a} & $<26$\tnote{a} & $< 78$\tnote{a} \\   
RX J0420.0-5022 &  $\geq $300\tnote{ass} & $<138$\tnote{b} & $<138$\tnote{b} \\ 

Geminga  &  $250^{+120 }_{-62}$ \tnote{c} & $142.2 \pm 1.2$\tnote{c} & $107.4 \pm 1.2$\tnote{c} \\

RX J0720.4-3125  & $361^{+172 }_{-88}$ \tnote{d} & $-93 \pm 1.2$\tnote{d} & $52.8 \pm 1.3$\tnote{d}\\

RX J0806.4-4123 &  240\tnote{ass} & $<76$\tnote{b} & $<76$\tnote{b}\\ 

RX J1856.6-3754 & $167^{+18 }_{-15}$ \tnote{d} & $326.7 \pm 0.8$ \tnote{e} & $-59.1 \pm 0.7$ \tnote{e} \\ 

PSR J1932+1059 & $361^{+10}_{-8}$ \tnote{f} & $94.03 \pm 0.14$\tnote{f}& $43.37 \pm 0.29$\tnote{f}\\ 

PSR J2124-3358\tnote{A} & $270 \pm 20$\tnote{g} & $-14.4 \pm 0.8$\tnote{h} & $-50  \pm 2$\tnote{h}\\ 

RBS 1774\tnote{K}& $\geq $300\tnote{ass} & n.k. & n.k.\\
\hline
\end{tabular}
\begin{tablenotes}
\item[K] 1RXS J214303.7+065419 
\item[A] PSR J2124-3358 turned out to be too old for our substellar companion
  search. See Online Appendix Sec.~\ref{psr2124} for a NIR limit of this NS star.
\item[a] \citet{Mignani2003}, $^b$  $2\sigma$ limit of \citet{Motch2006london},
\item[c] \citet{Faherty2007}, $^d$ \citet{Kaplan2007}, 
\item[e] \citet{Walter2002}, $^f$ \citet{Chatterjee2004}, 
\item[g] \citet{Gaensler2002},  $^h$ \citet{Hotan2006}
\item[{ass}] notes an assumed distance, which is a lower limit as indicated by estimations from X-ray measured hydrogen column density (e.g. \citealt{Posselt2007}), see also text. \protect\\
\end{tablenotes}
\end{threeparttable}
Individual images with poor seeing or high airmass have been excluded.
For comparison also ESO's ISAAC pipeline has been applied.
Multiple observations of the same object have been reduced individually. The results have then been combined using $IRAF$ \citep{Tody1986,Tody1993} if observing conditions were similar.
We fitted the absolute astrometry for the first epoch observation with the
2MASS point source catalogue (PSC, \citealt{Skrutskie2006}) using the \textsl{Graphical
Astronomy and Image Analysis Tool}  ($GAIA$) \citep{gaia}.
In case of many relatively bright 2MASS sources in the field (e.g. RX~J1856.6-3754), a positional accuracy of 300\,mas ($3\sigma$, \citealt{Skrutskie2006}) can be achieved. In some fields only few 2MASS sources are apparent. In these cases we used the  USNO B1 catalogue for which \citet{Monet2003} stated an astrometric uncertainty of 200 mas for absolute astrometry.
However, this optical catalogue is based on much older observations than 2MASS and its entries must not necessarily coincide with apparently matching NIR detections in our observations. 
We use absolute astrometry only to derive NIR limits for the neutron star
position taking an appropriate error circle to consider astrometric
calibration effects as well as proper motions of the neutron stars. 
Neutron star positions are preferably either a VLBI position
  (e.g. PSR~J1932+1059) or the position of the optical counterpart
  (e.g. RX~J1856.6-3754), if those are not known the (XMM-$Newton$) X-ray position (e.g. RBS
  1774) is taken, references are corresponding to those of the proper motions in
  table~\ref{objects} or those of the individual introductions in Sec.~\ref{gemingasec} to
  \ref{onlyfirst}. The NS's positional uncertainties have no impact on our
  search for substellar companions; in Sec.~\ref{limits} appropriate sky
  regions are considered for a possible NIR counterpart to a NS.
\label{onlyfirst}

No standard stars were explicitely observed for this project since only rough photometry was needed.
We applied for photometric calibration the zeropoints obtained by ESO during their regular $ISAAC$ calibration plan as starting point, and calibrated further  with 2MASS photometry by using -- preferably high quality -- 2MASS point sources in the field (for individual details see Sec.~\ref{result}). 

\subsection{Second epoch observations}
All the second epoch observations (see Tab.~\ref{obs}) were also carried out in the $H$-band in a similar jitter mode as the first epoch observations.
The second epoch observations of RX~J1856.6-3754 and PSR~J1932+1059 in May
2006 were moved by ESO to the 3.5\,m New Technology Telescope (NTT) equipped
with the {\it{Son of ISAAC}} instrument ($SOFI$) (instead of VLT $ISAAC$). 
The two fields were observed with $SOFI$'s Small field objective for around three to four hours during each of the four nights.
The seeing was usually better than $1\arcsec$, and during the second and third
night most of the time even better than $0.8\arcsec$, and in the last night
very variable. 
The observations were reduced by applying the $SOFI$ pipeline procedures at the telescope. We cross-checked the result by reducing individual runs also with $Eclipse$. The further data reduction procedure was the same as for the first epoch observations applying $IRAF$ and $GAIA$.
The second epoch $SOFI$ observations were up to 1~mag less deep than the $ISAAC$ observations.\\
In May 2007 VLT {\it{NAos COnica}} ($NACO$) observations were done for the inner field around PSR~J1932+1059 using the coronographic mode for a bright star close to the pulsar.
$NACO$ is an adaptive optical (AO) imager and spectrometer and has been operated with the S54 camera.
The Visual dichronic element and wavefront sensor (0.45 - 1.0 $\mu$m) and the opaque $1.4\arcsec$ coronograph were used.
A jitter box of $20\arcsec$, coronographic individual exposures of 60\,s, sky individual exposures of 0.3454\,s, 4 "object" offset positions and 6 "sky" offset positions  were further parameters of this $NACO$ run.
Observing conditions for the roughly one hour observation were very good with seeing better than 0.6 at an airmass of around 1.3 and a reached strehl ratio of around 17\,\% .
The $NACO$ observations have been reduced applying the $jitter$ engine of
$Eclipse$ (for more details, see Sec.~\ref{detailsobs} of the  Online Appendix).\\
Second epoch observations of RX~J0720.4-3125 and Geminga were obtained with VLT $ISAAC$ in  
December 2006 and January/February 2007 respectively (for more details see Tab.~\ref{obs}).
The data reduction procedure was the same as for the first epoch $ISAAC$
observations applying $Eclipse$, $ISAAC$ pipeline, $IRAF$ and $GAIA$.\\
The $NACO$ field of view is $56\arcsec \times 56\arcsec$ and therefore smaller
than those of $ISAAC$ and $SOFI$ with $154\arcsec \times 154\arcsec$. The
(nominal) pixel sizes are $0.054 \arcsec$ per pixel for NACO, $0.148 \arcsec$ per pixel
for $ISAAC$, and $0.144 \arcsec$ per pixel for $SOFI$.

\section{Method}
\label{method}
From the known proper motions of the neutron stars, it is possible to
calculate the expected shifts in pixel coordinates. First and second epoch
observations have to be compared regarding object detections, magnitudes and
positions. Due to the chosen time span it should be possible to distinguish
clearly the expected shifts for co-moving objects from position errors.  
We tested several source detection algorithms, e.g. the  {\textsl{Source
      Extractor}} by \citet{Bertin1996}, to determine the best   relative
  astrometry. 
  We refer to Sec.~\ref{detailsm} for more details. Here we concentrate on our
  final approach for the relative astrometry -- the \stf-code by
  \citet{Diolaiti2000SPIE}, an IDL-based point source detection algorithm for
crowded fields which starts by determining the point spread function (PSF)
from the science image for further fitting. The performance of \stf has been
tested, e.g., by \citet{Diolaiti2000} and \citet{Christou2004}. The latter
found, for example, that in fields with overlapping stellar PSFs \stf was
capable to produce consistent astrometry of the order of one tenth of the PSF
or better. Photometric errors are usually better than around 0.1\,mag
\citep{Diolaiti2000,Christou2004}. For very faint sources \stf is sensitive to
the structures in the determined model PSF. 
We carefully selected the reference sources for the model PSF and noise estimation.
Only objects detected with a SNR of at least three in both epochs - $ISAAC$,
$SOFI$ and $NACO$ observations - were considered for the search for co-moving
objects. 
The correlation factor for the \stf detections were usually better than 0.7. 
All sources have been manually inspected and apparent galaxies or artefacts have been excluded.
If we compared sources detected by two different instruments, then the
($SOFI$, $NACO$) pixel coordinates were converted to coordinates within the
$ISAAC$ reference frame.
Source lists of two epochs were then correlated and we derived the positional
shifts as presented in Section~\ref{result}.
In case of two epochs of $ISAAC$ observations, we minimized systematic
instrument-related errors like field distortions by a similar observation
setup.
In case of different instruments/telescopes, the additional systematic errors
showed up in a broader scattering of the computed positional shifts.  
In case of the $NACO$ observations this extra error is negligible in the chosen $ISAAC$
reference frame.

 If no second epoch observation was obtained, e.g., due to still unknown proper
motion, we investigated the NIR sources in a 5000 AU circle around the NS position
for their optical-NIR colors to identify potentially interesting objects for
follow-up observations. We refer to Sec.~\ref{detailsno2} for more details on
the procedure. 

At the positions of the NSs, we checked also for faint NIR-objects which could
be unexpected possible counterparts of the NSs themselves. We determined an
upper $H$-band limit for the NS if there is no NIR object within the
positional error by taking the faintest, closeby object having a SNR larger
than 3.
We checked, whether we could reach deeper NIR-limits at the position of the
NSs by combining first and second epoch observations.

\section{Results}
\label{result}
\subsection{Pixel shifts between two epochs of fast moving neutron stars}
\begin{table*}
\caption[Second Epoch Parameters]{
For objects with second epoch observations the following parameters are listed
in this table: the brightest $H$-band magnitudes expected for a brown dwarf companion (m$_{BD}$) or a
gas planet of 13 Jupiter mass (m$_{13Jup}$) taking into account the NS's respective distance and
age (see text and e.g. table~\ref{objects}). 
Furthermore, the reached magnitude limit( m$_{lim}$) in both epochs, the time difference
between first and second epoch, the covered projected physical separations (r$_{proj}$), and the pixel scale of the reference $ISAAC$ image
(refX, refY) are listed.
Reference image axes are roughly aligned with RA (X) and DEC (Y) within 1 degree, however
sometimes with opposite direction (indicated by '-' instead of '+').   
}
\label{secpar}

\begin {tabular}{@{} l | c  c  c  |  c  c | c  c  c  c @{}}
\hline
Neutron Star & m$_{BD}$ & m$_{13Jup}$ &  m$_{lim}$ & epoch diff. &
r$_{proj}$ & refX & refY & align$_X$ & align$_Y$ \\
             & [mag] & [mag] & [mag] & [yr] & [AU] & [mas] & [mas] & +/- & +/-\\
\hline
Geminga & 18.2 & 22.3 & 22.4 & 3.05 & 10000 & 148.4 & 148.4  & + & {-} \\

RX~J0720.4-3125 &  19.0 & 23.0 & 22.9 & 2.95 & 20000 & 147.7 & 147.7 &+ & - \\

RX~J1856.6-3754 & 17.3 & 21.4 & 21.5 & 3.0 & 10800 & 147.8 & 147.7 & + &+ \\

PSR~J1932+1059 & 21.5 & 23.5 & 21.5  &2.9, 3.9 &  see text & 147.5 & 147.3  & - &+\\
\hline
\end{tabular}
\end{table*}

\begin{figure*}
\center
\begin{minipage}[l]{.32\textwidth}
 \hspace{-1.5cm}
 \psfig{figure=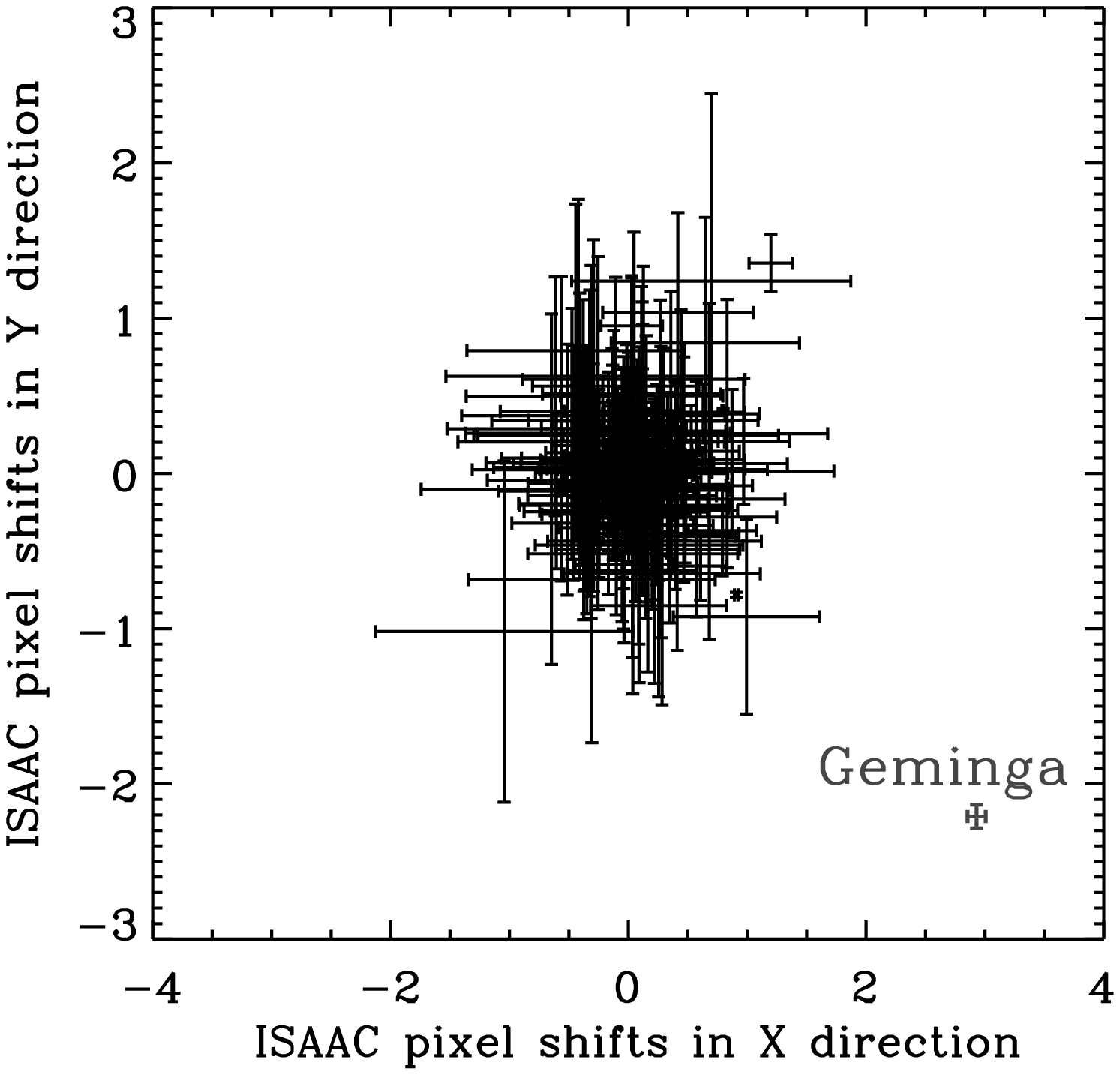,width=7cm}
\end{minipage}
\begin{minipage}[l]{.32\textwidth}
 \hspace{-1cm}
 \psfig{figure=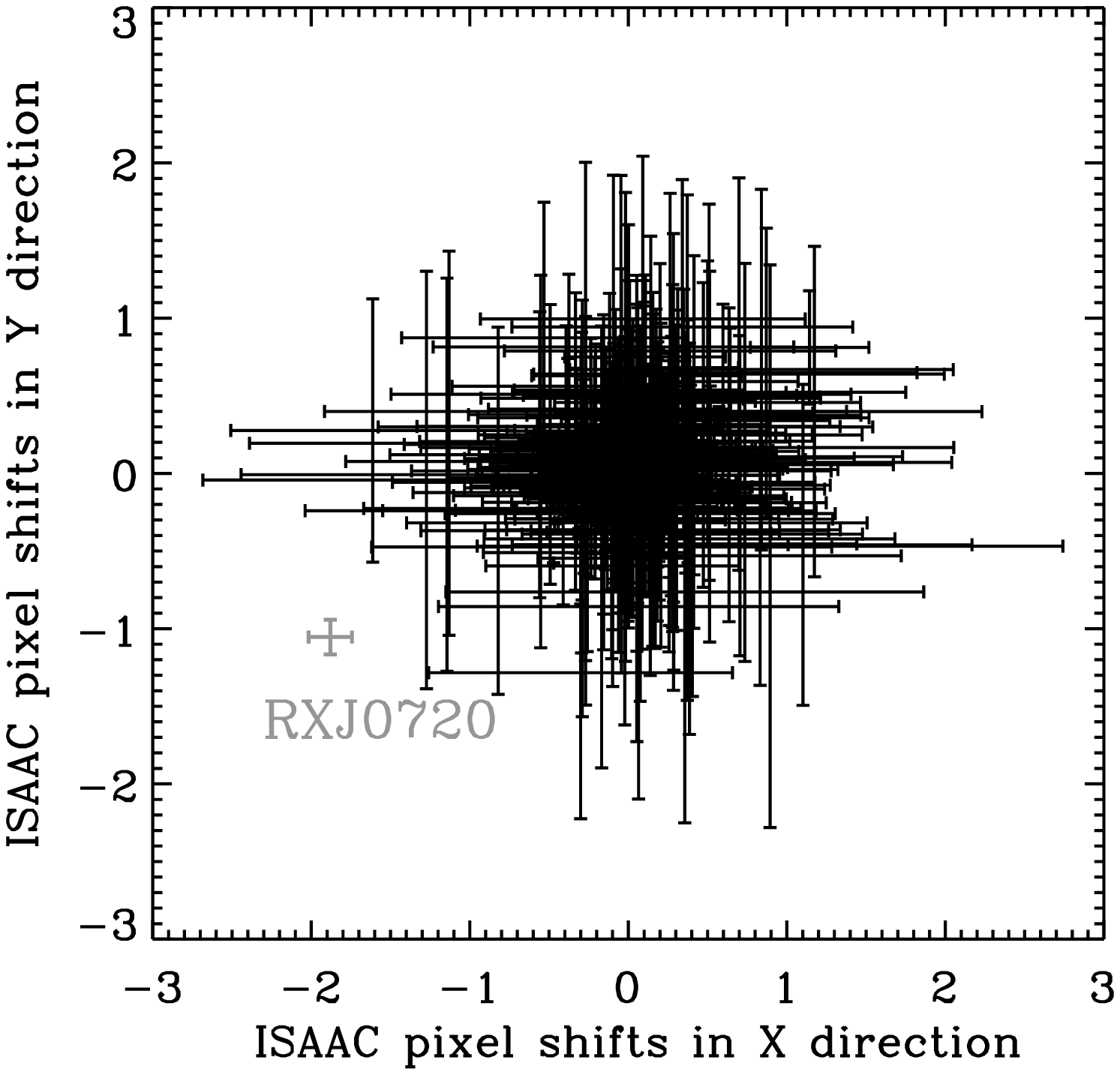,width=7cm}
\end{minipage}
\begin{minipage}[l]{.32\textwidth}
 \hspace{-0.5cm}
 \psfig{figure=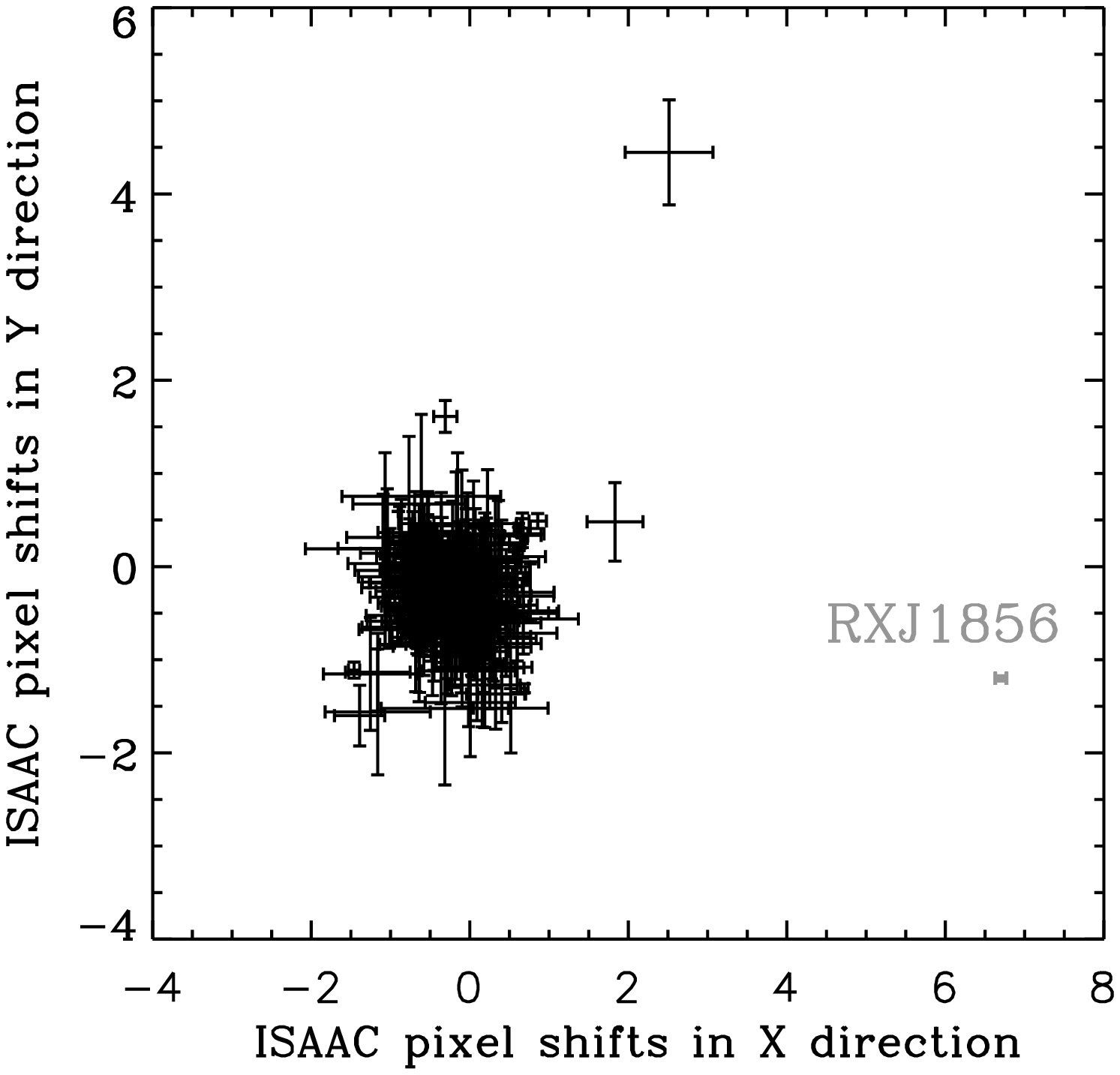,width=7cm}
\end{minipage}
 \begin{flushleft}
 \caption{\textbf{Pixel shifts between the first and second epoch NIR observations of Geminga,
RX J0720.4-3125, RX J1856.6-3754} (from left to right).}
Errors are $3\sigma$-\stf positional errors. 
The grey positions marked with Geminga, RXJ0720, RXJ1856 show the positions a
possible co-moving object would have. 
For more details regarding the $ISAAC$ reference frame pixel scale, axes
orientation and covered magnitudes see Tab.~\ref{secpar} as well as 
Sec.~\ref{gemingasec} to \ref{1856sec}.
 \label{fig:shiftsthree}  
 \end{flushleft}
\endcenter
\end{figure*}

\label{res1}
All faint NIR sources near the known NS position were analysed as outlined in Sec.~\ref{method}.
Co-moving companion candidates, found by comparison of first and second epoch observations, would have a high probability to be true companions, because NSs have
very large proper motions, and it would be unlikely
that an unrelated, distant, faint background object 
would share such a large proper motion.

Table~\ref{secpar} lists some parameters of the analysis. 
Additional notes on these and the results can be found in Sec.~\ref{gemingasec} to \ref{1932sec}. 

\subsubsection{Geminga}
\label{gemingasec}
Geminga is the proto-type of the $\gamma$-ray pulsars (see e.g. \citealt{Bignami1996}). The distance of Geminga is still under discussion. \citet{Faherty2007} recently reported a newly estimated parallactic distance of $250^{+120 }_{-62}$~pc in comparison  with the older value of $157^{+59}_{-34}$~pc by \citet{Caraveo1996,Caraveo1998}.
We adopt in the following the most recent distance by \citet{Faherty2007}.
Geminga has a spin-down age of around 340000 years. 
The lowest age considered by B97 (or similar papers, e.g.,
\citealt{Baraffe1998, Baraffe2002}) is 1~Myr. 
We consider 383 sources in the whole $ISAAC$ field of Geminga, most of them fainter than 18.2~mag.
The pixel shifts of the objects, common in both epochs, are plotted in Fig.~\ref{fig:shiftsthree}.
There is no apparent co-moving source within $\approx 10 000$~ AU around Geminga down to $H=22.4$~mag.
For a 1~Myr age (see above) and 250~pc this corresponds to exclusion of all young companions having more than 
12 Jupiter masses following the calculations by B97.\\

\subsubsection{RX~J0720.4-3125}
\label{0720sec}
RX J0720.4-3125 is the second brightest of the X-ray thermal isolated neutron stars.
We assume in the following a similar age -- 1~Myr -- as for RX J1856.6-3754.  
The distance to RX J0720.4-3125 has still a large uncertainty. \citet{Kaplan2007} reported a parallactic distance from optical observations to be $361^{+172 }_{-88}$~pc. 
The hydrogen column density fitted in the X-ray spectrum indicates a distance of 240 to 270 pc taking into account the local inhomogenously distributed interstellar medium \citep{Posselt2007}. 
In the second epoch, we had to reduce the correlation coefficient constraint to
0.65 to re-identify all first epoch detections. 14 sources (out of the
considered 581) in the second epoch have correlation coefficients between 0.65
and 0.7.
The pixel shifts of the objects common in both epochs are plotted in Fig.~\ref{fig:shiftsthree}.
There is no apparent co-moving source within $\approx 20 000$~ AU around RX~J0720.4-3125 down to $H=22.9$~mag.
For a 1~Myr age and 361~pc distance, this corresponds to exclusion of all young
companions having more than 15 Jupiter masses following the calculations by
B97.\\

\subsubsection{RX~J1856.6-3754}
\label{1856sec}
RX J1856.6-3754 is the brightest of the X-ray thermal neutron stars.
Its $HST$-determined parallax of $117 \pm 12$~pc by \citet{Walter2002}, was
revised to $142^{+58}_{-39}$~pc by \citet{Kaplan2002} and recently further
revised to preliminary $167^{+18 }_{-15}$~pc \citep{Kaplan2007}. We adopt in the following this most recent value.
The age of RX J1856.6-3754 is assumed to be around 1~Myr given the proper motion and the likely birth place in Upp Sco (e.g. \citealt{Walter2002}).
There are 525 common objects having a SNR larger than three in both, the $ISAAC$ and
$SOFI$ observations, 448 of them fainter than $H=17.3$~mag, and two with $H
\geq 21.4$~mag.
In Fig.~\ref{fig:shiftsthree} we plot the pixel shifts between the two epochs as well as the pixel shift of the neutron star which would be the same for a co-moving object.
We did not account for pixel scale errors or instrumental distortion
errors. Both are present in the scatter distribution of the object's pixel
shifts.
We do not expect an additional positional error larger than one 
fifth pixel (mean $3\sigma$ positional uncertainty) due to coordinate
transformation ($SOFI$ to $ISAAC$ reference frame). 
Even accounting for this does not change the picture of Fig.~\ref{fig:shiftsthree}
-- there is no co-moving $H$-band source with SNR $\geq 3$ present in our
observations.
Applying the calculations of B97 for an assumed age of 1~Myr and a
distance of 167~pc, substellar companions heavier than 0.01 M$_\odot$ or 10.5
Jupiter masses can be excluded around RX J1856.6-3754.

A short note about the two objects with exceptional pixel shifts in the right-hand panel of Fig.~\ref{fig:shiftsthree}: 
The object with offset of about 1.8, 0.5 belongs to a triple $ISAAC$
source. It has an $ISAAC$ $H$-band magnitude of 20.6~mag, the other two
sources have  $H=19.6$~mag and  21.3~mag and are located roughly $1.2\arcsec$
from each other. In the $SOFI$ observations, these three sources are more
blurred and \stf can detect only two of them with SNR$\geq 3$ and correlation
coefficient $\geq 0.7$.
The derived second epoch position of the object in question is slightly
shifted towards the position of the not detected fainter object.
There were around ten such cases of non-detected multiple sources in this
particular field. Usually, the positions derived by \stf are only marginally
influenced. 
The pixel shift of around 2.5, 4.4 belongs to an 21.0 mag object which is
separated around $1.6 \arcsec$ from a 16.5~mag object. By eye inspection this
source is barely visible in the $SOFI$ observation. \stf manages to still
detect the faint source, however, the position seems not to be accurate for
this high contrast. 

\subsubsection{PSR~J1932+1059}
\label{1932sec}
\begin{figure*}
\center
\begin{minipage}[l]{.45\textwidth}
 \hspace{-1.5cm}
 \psfig{figure=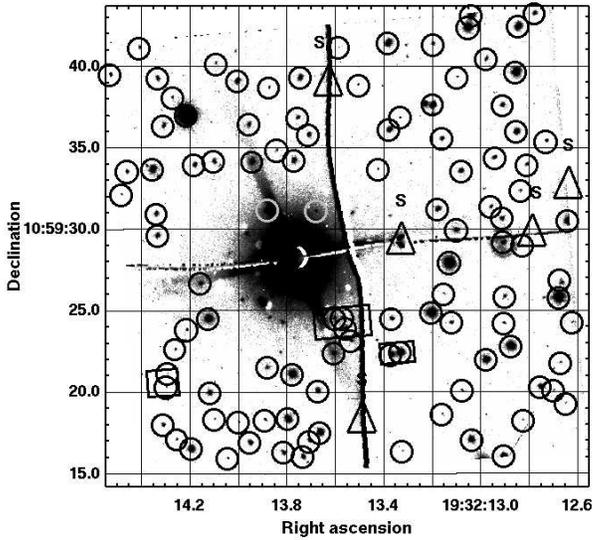,width=8cm}
\end{minipage}
\begin{minipage}[l]{.45\textwidth}
 \psfig{figure=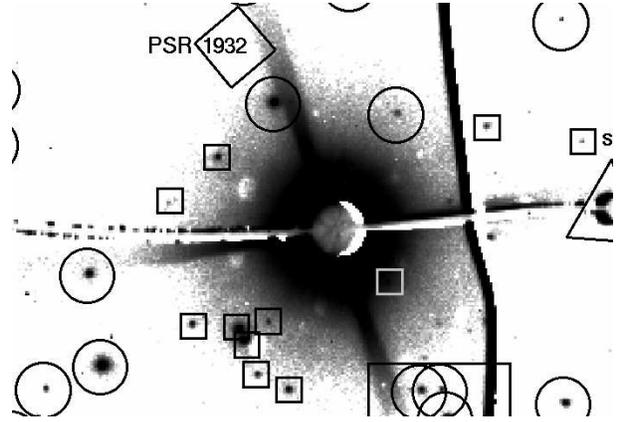,width=8cm}
\end{minipage}
 \begin{flushleft}
\vspace{-0.5cm}
 \caption{\textbf{The $NACO$-field of PSR J1932+1059} with circles indicating
   common sources in $NACO$ and $ISAAC$ observations. The big boxes are
   pointing toward multiple $NACO$ detections which have been identified as
   only one source in the $ISAAC$ image. Triangles with 'S' mark positions of
   sources for which we used the $SOFI$ observation for the common proper
   motion search.
   The right figure is a zoomed-in (roughly $14\arcsec \times 10\arcsec$) variant.
Sources not considered for the common proper motion search are marked with
squares. They are detected only in the $NACO$ image due to the coronographic
mode, but not in the $ISAAC$ observation. The (empty) position of the pulsar
is indicated by the diamond shape with $1\arcsec$ radius. \vspace{-0.5cm} }
 \label{field1932}  
 \end{flushleft}
\endcenter
\end{figure*}
PSR J1932+1059 is a radio pulsar. New VLBI measurements put this neutron star
at a distance of $361 \pm  10$ pc \citep{Chatterjee2004}, further away than
previously thought. Therefore, only brown dwarfs could be detected as possible
substellar companions given its spin down age of $3.1 \cdot 10^{6}$ years
\citep{ATNF2004}, and following the models by B97. 
PSR~J1932+1059 lies unfortunately in a crowded field with a very bright,
$H=8.8$~mag, object only $ \approx 2.5\arcsec$ away from the pulsar.
The combined $SOFI$ observations of May 2006 were more blurred than the $ISAAC$ ones due to the observing conditions (small visibility time window, not so good weather on the last night). The bright NIR object causes large "doughnut"-shaped artefacts (only in the $SOFI$ image) due to the jittering process.
While the \stf-Code is able to still detect reasonable well the NIR sources,
we avoided most of the artefacts and the bright source by using the $NACO$
coronographic observations from May 2007 instead. 
Note that the $NACO$ S54 camera has a pixel scale of around 54 mas comparing
to around 148 mas for $ISAAC$. $NACO$ \stf positional errors are much smaller
than positional errors for $ISAAC$.
NIR sources in an area of around $10000$~AU projected physical separation of the pulsar (at an assumed distance of 361~pc) have been considered for further analysis.
The area covered by $NACO$ extends in the east and north up to about 3000~AU,
and in the west and south up to roughly 6500~AU projected physical separations
from the NS. In this area, we used the $SOFI$ image only for four sources,
which we could not detect well in the $NACO$ image (wire construction
shadow). 
Due to the coronographic $NACO$ AO mode and smaller pixel scale there were more sources detected as in the $ISAAC$ image.
By checking the NIR sources found in both epochs, several single $ISAAC$ sources turned out to be actually multiple $NACO$ sources. \stf positional errors for the $ISAAC$ sources are not conspicuously high in these cases indicating the limits of the \stf code. 
In case of a relatively bright main component in the multiple source, it is
reasonable to assume that this specific source was actually detected in the
$ISAAC$ image. 
If the multiple $NACO$ sources had nearly the same magnitudes we compared the
flux-weighted average position of the individual $NACO$ sources to the one
$ISAAC$ position in such cases. Furthermore, we looked at the pixel shift of
each individual $NACO$ source comparing to the one $ISAAC$ source aiming to
get a feeling how large pixel shifts due to unidentified multiple sources
could be.
All considered common $NACO$-$ISAAC$ sources are marked in Fig.~\ref{field1932}. 
Close ($< \approx 4 \arcsec$) to the bright star there are eleven $NACO$ sources which were not detected with $ISAAC$ at all. These sources have $H$-band magnitudes of around 18 to 21~mag. They are marked in Fig.~\ref{field1932} and have to be investigated for co-motion in another epoch.
   
\begin{figure*}
\center
\begin{minipage}[l]{.32\textwidth}
 \hspace{-1.5cm}
 \psfig{figure=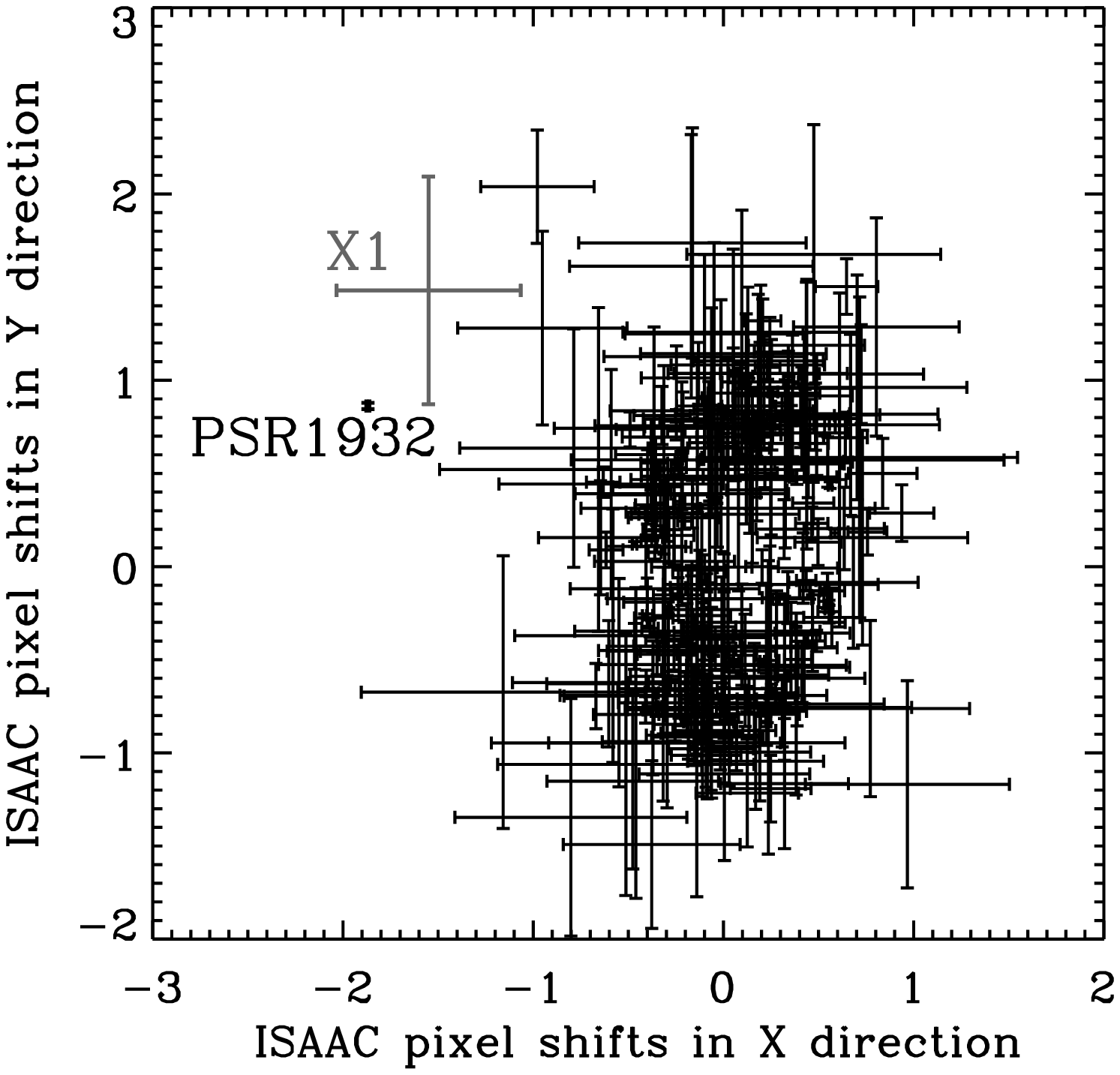,width=7cm}
\end{minipage}
\begin{minipage}[l]{.32\textwidth}
 \hspace{-1cm}
        \psfig{figure=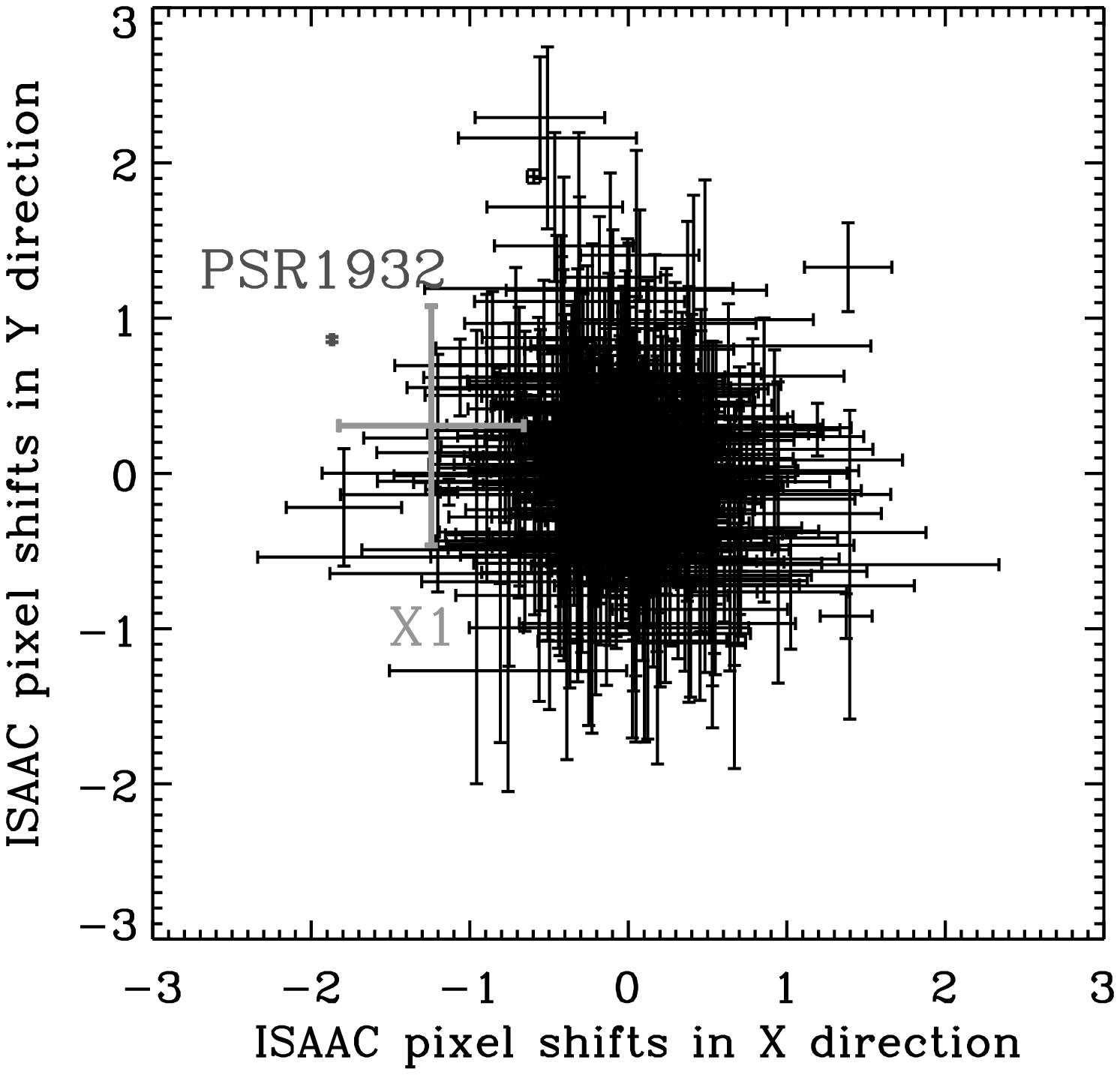,width=7cm}
\end{minipage}
\begin{minipage}[l]{.32\textwidth}
 \hspace{-0.5cm}
 \psfig{figure=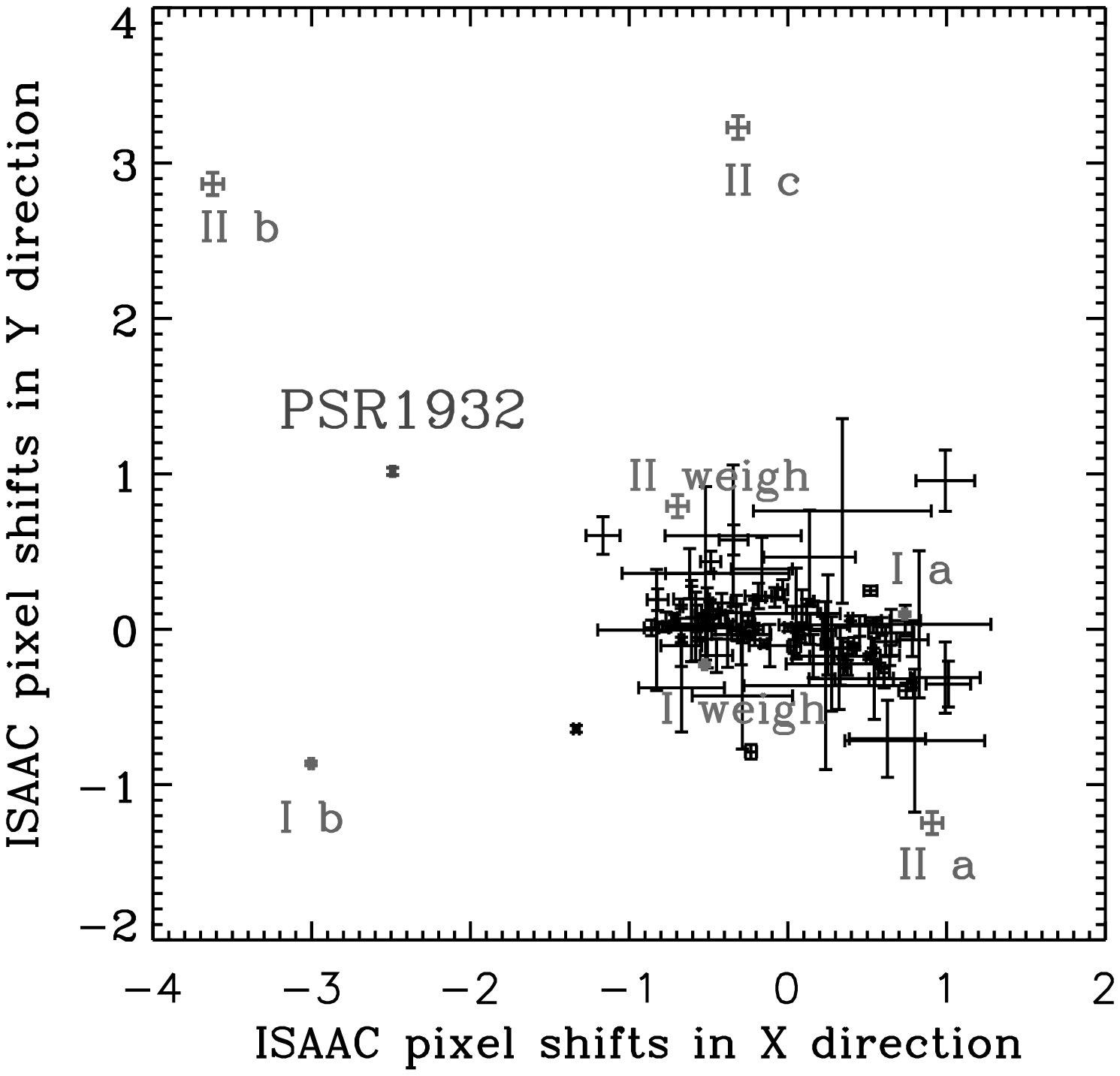,width=7cm}
\end{minipage}
 \begin{flushleft}
 \caption{\textbf{Pixel shifts between the first and second epoch NIR
     observations of PSR J1932+1059 }\protect\\ 
Errors are $3\sigma$-\stf positional errors. 
'PSR1932' marks the position of an object which would co-move with PSR
J1932+1059, the shown error bars correspond to the $3\sigma$ uncertainty of
the pulsars proper motion. 
{\bf The left plot} is for a distortion-{\emph{un}}corrected analysis of the
$ISAAC$ and $SOFI$ images.
The 21.2~mag object labeled 'X1' is $25.6\arcsec$ north-east of the pulsar.
{\bf The middle plot} shows the pixel shifts of the same objects, but for distortion-corrected $ISAAC$ and $SOFI$ images. 
The re-sampled $ISAAC$ pixel scale is 148.4 mas for $x$, and 148.4 mas in $y$.
The object X1 and the pulsars proper motion do marginally $not$ coincide within their $3\sigma$ uncertainties.
Pixel shifts for common sources in the $ISAAC$ and the $NACO$ observation are
shown in the {\bf right plot}.
As described in the text single $ISAAC$ sources turned out to be multiple
$NACO$ sources. Pixel shifts of the two components of the  $ISAAC$ source 'I'
and the three components of the  $ISAAC$ source 'II' are labeled in the plot
as well as the flux-weighted average positions of Ia,b and IIa-c. \hspace{0.5cm}
See Tab.~\ref{secpar} and text for more details. \vspace{-0.5cm}}
 \label{shifts1932}  
 \end{flushleft}
\endcenter
\end{figure*}

The pixel shifts of the 284 sources considered for the common proper motion search in the $SOFI$ and the $ISAAC$ observations are plotted in Fig.~\ref{shifts1932}.
Due to the observing conditions the errors in the $SOFI$ observations are relatively large. 
There is a faint 21.2~mag object labeled 'X1' in Fig.~\ref{shifts1932} (left) whose $3\sigma$ error bars overlap with the position of an object which would share common proper motion with PSR J1932+1059.
The spread in the pixel shifts in Fig.~\ref{shifts1932}, especially the fragmentation in two 'clouds' indicate systematic effects. The $SOFI$ small field is known for strong distortions in some parts of the instruments field of view.
This may explain X1's position in the pixel shift plot. To account for distortions we applied {\textsl{SCAMP}} \citep{Bertin2006scamp} with a polynomial degree of five for both, $ISAAC$ and $SOFI$ images. As expected there are large $SOFI$ distortions while for $ISAAC$ they are small. We applied then {\textsl{SWarp}} \footnote[1]{http://terapix.iap.fr/rubrique.php?id\_rubrique=49} 
by E. Bertin (Terapix project) for resampling and proceeded with \stf for a similar analysis as described above.
The resulting pixel shift plot is shown in Fig.~\ref{shifts1932} (center). The sources are much more clustered in this figure. Sources with large pixel shifts are either indeed moving objects or they are close to bright stars with their detection parameters possibly slightly affected.
The X1-object and  the pulsars proper motion do not coincide within their $3\sigma$ uncertainties, however only marginally. It is very likely  that X1 is a background object.
If unexpectedly X1 is a co-moving companion, this source would be at around 9200~AU projected physical separation from the pulsar, and applying the calculations by B97, it would have a mass of around 42 Jupiter masses for a distance of 361~pc and an age of $3.1 \cdot 10^{6}$ years.
There is no known object within $10\arcsec$ of X1 in the SIMBAD-database.
Unfortunately, X1 was not in the field of view of our $NACO$ observations. 

The pixel shifts of the 104 $NACO$ sources with respect to the earlier $ISAAC$ observations are shown in Fig.~\ref{shifts1932}. 
The $NACO$ sources 'I a' and 'I b' belong to one 17.2~mag $ISAAC$ detection as well as 'II a' - 'II c' to another 19.1~mag one. 
Here, it seems likely that the single $ISAAC$ sources are actually a merger of the multiple $NACO$ ones. 
The flux-weighted average $NACO$ positions can be regarded as the putative $ISAAC$ positions if this instrument would have been used. 
The respective flux-weighted average pixel shifts 'I weigh' and 'II weigh' are situated within the bulk of the pixel shifts of the other sources.  Therefore, the $ISAAC$ sources 'I' and 'II' can be ruled out as co-moving objects.
The large pixel shifts of I a-b, II a-c (each $NACO$ source considered separately) illustrate that analysing detections by two instruments with different spatial resolution is of limited use given the high precision needed for the relative astrometry.
Thus, to sufficiently probe the co-motion of all individual $NACO$ sources another $NACO$ epoch would be necessary. 
Overall, Fig.~\ref{shifts1932}  shows for the common $NACO$-$ISAAC$ sources that there is no pixel shift which would be expected from a co-moving object.
However, another $NACO$ observation would be needed to confirm this.

The current state of our search around PSR J1932+1054 is that we found no co-moving object down to $H=21.5$~mag comparing the $ISAAC$ with the $SOFI$ observation, as well as the $ISAAC$ with the $NACO$ observation. 
At the given distance and age, there are probably no substellar companions down to 42 Jupiter masses. 
However, this result has to be seen in the light of the described uncertainties due to the different instruments and the close bright star. Further investigations are needed for confirmation. 

\subsection{Neutron stars with only first epoch observations}
In the following, the first epoch $H$-band observations in connection with
available optical data are discussed to judge about the necessity of second
epochs to the $ISAAC$ observations. 
   
\begin{figure*}
\center
\begin{minipage}[l]{.45\textwidth}
 \hspace{-1.0cm}
 \psfig{figure=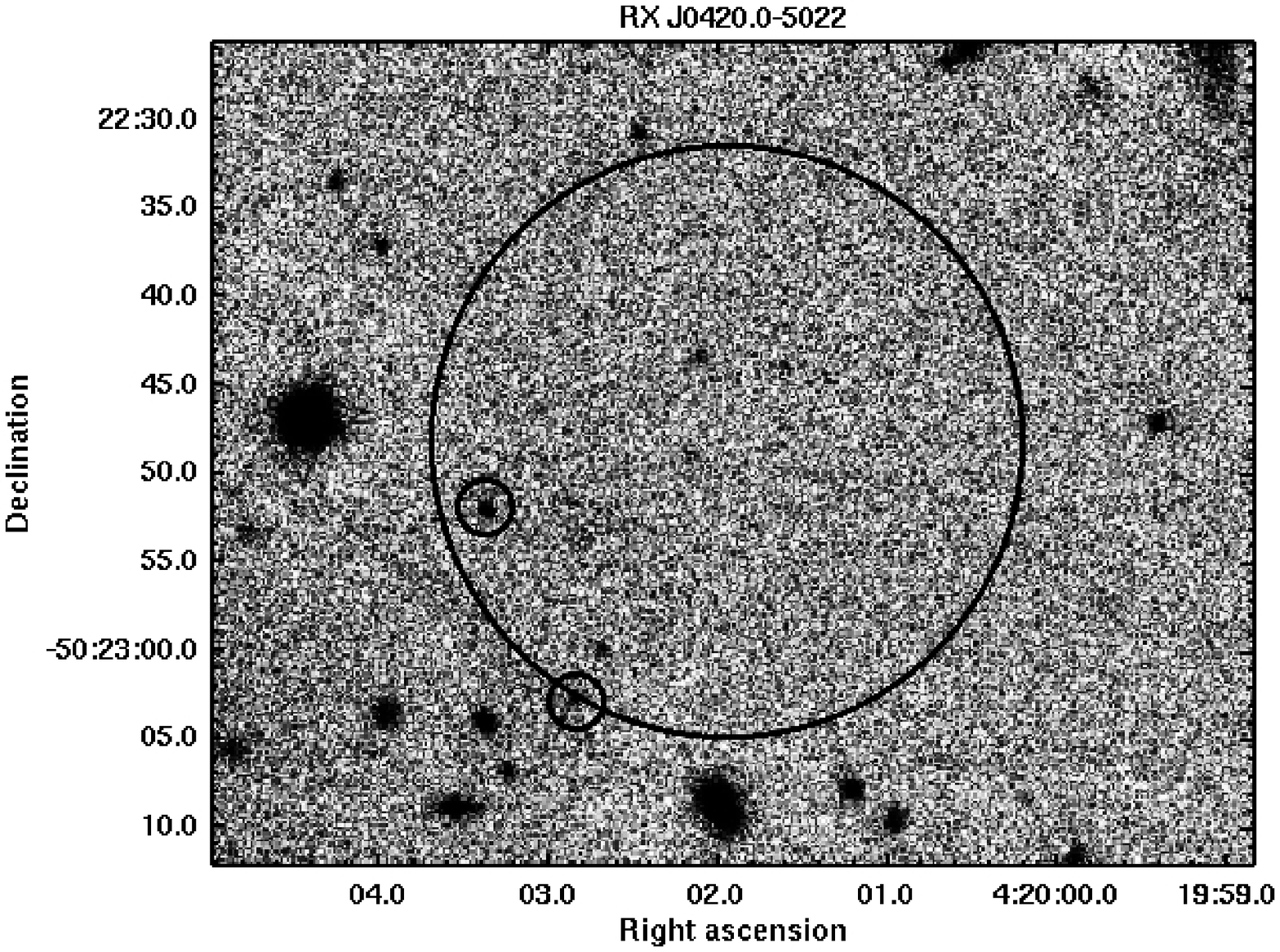,width=8cm}
\end{minipage}
\begin{minipage}[l]{.45\textwidth}
 \hspace{-0.5cm}
 \psfig{figure=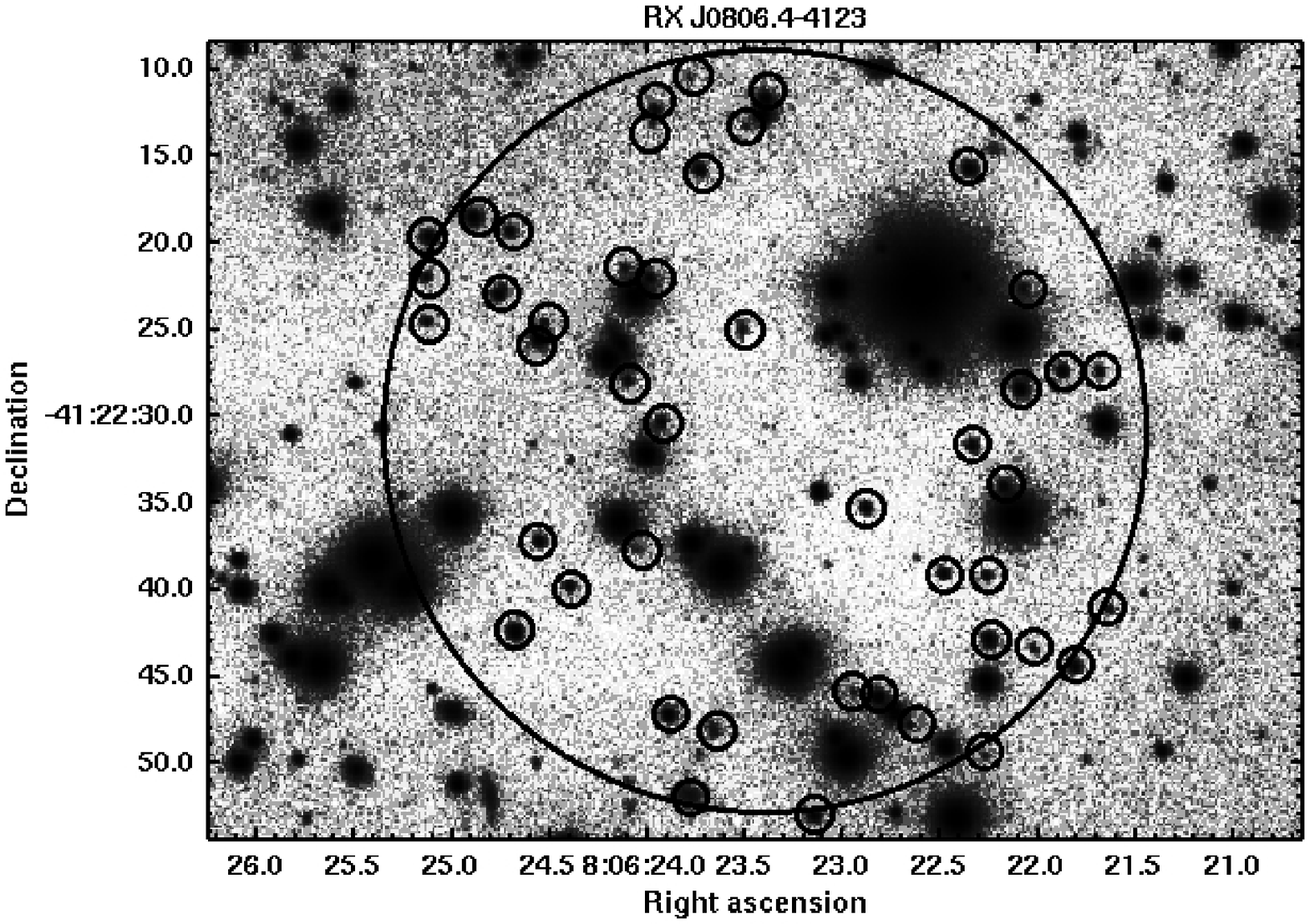,width=8.5cm}
\end{minipage}
\vspace{-0.5cm}
 \begin{flushleft}
 \caption{\textbf{Faint NIR-objects around RX J0420.0-5022 and RX J0806.4-4123} which may be interesting for second epoch observations. 
The big circles have a radius of 5000~AU at the respective distances. The
marked faint objects have a detection SNR of at least three. 
\vspace{-0.7cm}} 
\label{fig:}  
 \end{flushleft}
\endcenter
\end{figure*}

\paragraph{PSR J0108-1431} is one of the closest radio pulsars
according to its very low dispersion measure ($2.38 \pm 0.01$~cm$^{-3}$ pc,
\citealt{Amico98}). The faint pulsar has a distance of 130~pc and seems to be
older (170 Myrs, \citealt{Mignani2003}) than we assumed initially. 
Possible brown dwarfs around PSR J0108-1431 are expected to be fainter than  $H=20.8$\,mag.
The obtained photometric magnitudes for the objects of interest have large error bars due to the availability of only two bright 2MASS sources for the zeropoint determination. Astrometric positions were refined using USNO B1 point source positions.
Within 5000~AU there is only one source fainter than $H=20.8$\,mag having $H=21.0$\,mag. 
We obtained the archival VLT $FORS1$ images in the V-band, reduced the data applying bias subtraction and flat-fielding and calibrated the source $V$-band magnitudes with the tabulated values of \citet{Mignani2003}. The faint NIR-object is detected in the $V-band$ with 25.6~mag. Therefore, the $V-H$-color is 4.6~mag. This color is typical for spectral type M2~V or M3~V according to \citet{Kenyon1995}. Even when taking an error of 1~mag into account the source is no candidate for a substellar companion around PSR~J0108-1431. Thus, there are no substellar companion candidates for second epoch observations from our VLT $ISAAC$ image within 5000~AU around PSR~J0108-1431.

\paragraph{RX J0420.0-5022} is one of the X-ray thermal
isolated neutron stars (e.g. \citealt{Haberl2004}). Neither distance nor
proper motion were measured due to a still insecure optical counterpart. The
hydrogen column density fitted in the X-ray spectrum indicates a distance of
very roughly 300~pc. Assuming a similar age as that of the better known XTINS
RX J1856.6-3754, one would expect substellar companions to be fainter than
$H=19$\,mag at this distance. 
Only two objects are present within 5000~AU around the neutron star position for an assumed distance of 300~pc. 
There are optical archival observations with ESO's VLT equipped with $FORS1$
(66.D-0128). The three $B$-band observations were added and the photometry was
calibrated with the USNO B1 catalogue due to missing archival
calibrations. 
At the position of the two interesting sources there are no optical sources
down to the $\approx 26$~mag limit of  the $FORS1$ $B$-band observations. 
Thus, the NIR-sources could not be excluded as possible substellar companions by their $B-H$ color.
Only the second epoch observation, as soon as the proper motion is constrained, can clarify possible companionship.

\paragraph{RX J0806.4-4123} is one of the faintest of the seven XTINSs (e.g. \citealt{Haberl2004}). Neither proper motion nor distance were measured due to a still missing optical counterpart. The hydrogen column density fitted in the X-ray spectrum indicates a distance of $240 \pm 25$~pc taking into account the local imhomgenously distributed ISM \citep{Posselt2007}. 
Possible substellar companions around RX J0806.4-4123 should have $H$-band
magnitudes fainter than 18.6\,mag
We find around 50 interesting NIR-sources within 5000~AU down to $H=22.0$~mag having a minimal SNR of three.
We observed the very crowded field in the $V$-band using ESO's NTT equipped
with the $Superb$ $Seeing$ $Imager - 2$ ($SUSI2$) in December 2004. The data
were reduced applying dark field subtraction and flat fielding. Bias
subtraction was neglected. Astrometry of the optical observation was obtained
by fitting with the USNO B1 point source positions. The photometry was
obtained by a standard field observation reduced in the same way as the
science field. The derived limit for $3\sigma$ detections is $V=24.8$~mag.
A few NIR-sources could be excluded as possible substellar companions due to
their low $V-H$-color. However, most of the interesting objects are not
detected in the $V$-band. All the final 45 objects will have to be
investigated in the second-epoch observation.

\begin{figure}[t]
 \hspace{-0.1cm}
 \psfig{figure=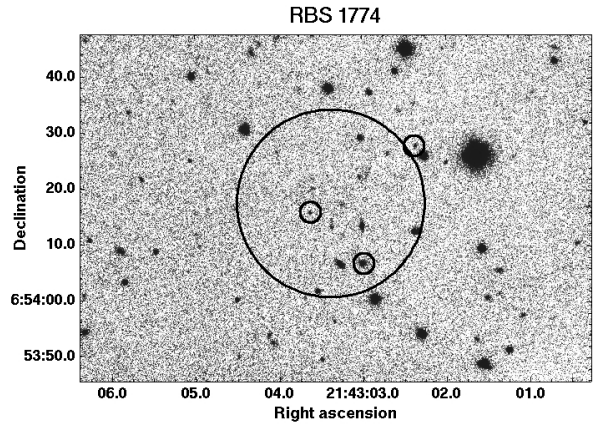,width=8cm}
      \caption{Faint NIR-objects around RXS J214303.7+065419 which may be interesting for second epoch observations. The big circle would have a radius of 5000~AU at 300~pc. 
The marked faint objects have a detection SNR of at least three.\vspace{-0.3cm} }
 \label{rxj2143}
\end{figure}

\paragraph{RXS J214303.7+065419,} also called  RBS 1774, is a XTINS. Currently, we do not know of an optical counterpart down to $r\arcmin=25.7$~mag \citep{Rea2007}. 
The most recent X-ray position by $Chandra$ has an uncertainty radius of
$0.6\arcsec$ \citep{Rea2007}. 
Here, we assume as a lower distance limit 300~pc and an age similar to that of
RX J1856.6-3754. Substellar companions are then expected to be fainter than
$H=19$\,mag. 
There are nine interesting objects within 5000~AU around the neutron star
position. 
Due to the availability of only three relatively bright and clustered 2MASS
point sources the photometric error of the sources is around $1$~mag,
especially for faint objects. The 2MASS obtained astrometry was improved by
using additional USNO B1 point source positions.
The archival ESO $EMMI$ $R$-band observations by \cite{Zampieri2001} have a
limit of $R=23$~mag. $EMMI$ detected objects, some also present as USNO B1
point sources, provided the $R-H$ color for the NIR-sources. For five
NIR-sources only the limit of $R=23$~mag could be applied and thus no
statement of a color in favour for substellar companionship is possible. The
four sources with a known $R$-band magnitude can be excluded as possible
substellar companions due to their $R-H$ color, even considering a color
magnitude error of 2~mag.
Two of the sources are elongated and are probably galaxies.
Whether the remaining three sources are co-moving objects can only be
constrained with the knowledge of the still unknown proper motion.

\subsection{H-band limits at the neutron star positions}
\label{limits}

\begin{table*}
\begin{minipage}[t]{19cm}
\caption{\textbf{$H$-band and mass limits obtained from the observations} \protect\\  
Listed are for each object: $H_{lim}$ indicates the faintest $H$-magnitude with a detection SNR of three, which one could measure at the position of the neutron star. The $1 \sigma$ error is a result of the \stf photometric error and the zeropoint error and an estimated/assumed error from comparison with 2MASS. See also text. $ZP$ states the magnitude of the zeropoint, which were obtained in the frame of ESO's calibration plan. $^{e2}$ indicates that two epochs have been used. $SE-2MASS-limits$ lists the {\textsl{Source Extractor}} limits if using relative 2MASS photometry. The assigned error indicates poor or high number of available, suited 2MASS PSC sources, but includes also detection routine uncertainties.    
For comparison, the next column lists the values by \citet{Curto2007}, which they derived from our observations.
They stated a general magnitude error of 0.1 to 0.2 mag.
The last column notes the reached mass limit (from Sec.~\ref{res1}), $\cdots$ notes that the found objects are interesting for a second epoch observation, -- states the opposite.
}
\label{hlimtab}
\renewcommand{\footnoterule}{}  
\centering
\begin {tabular}{@{} l   c  c  c  c | c @{}}
\hline
Object & $H_{lim} \pm 1\sigma$ & $ZP$ & $SE-2MASS-limits$ & \citet{Curto2007} & Mass-limit\\
       &  [mag]    &   [mag]    & [mag]  & [mag] & [Jupiter masses]\\
       &                  &         &        &   & \\
\hline

PSR J0108-1431  & $21.42 \pm 0.16$ & 24.66  & $21.5 \pm 1.5$ & $\cdots$ & --\\

RX J0420.0-5022 & $21.84 \pm 0.16$ & 24.61  & $21.5 \pm 0.6$  & 21.7 & $\cdots$\\

Geminga\footnote{Geminga has been detected in a deeper HST $F160W$ observation
with roughly 24.1\,mag \citep{Shibanov2006}}         & $22.66 \pm 0.14$ & 24.66$^{e2}$  & $22.0 \pm 0.4$ & $\cdots$ &12\\

RX J0720.4-3125 & $23.07 \pm 0.11$ & 24.63$^{e2}$  & $22.3  \pm 0.4$ & 22.7 & 15\\

RX J0806.4-4123 & $22.64 \pm 0.16$ & 24.59 & $22.6\pm 0.4$  & 22.9& $\cdots$ \\

RX J1856.6-3754 & $21.54 \pm 0.24$ & 24.65$^{e2}$ & $22.1 \pm 0.4$  & 21.5 & 11\\

PSR J1932+1059  & $20.31 \pm 0.18$ & 23.49 & $\cdots$  & $\cdots$ & 42\\ 

RBS 1774        & $22.04 \pm 0.16$ & 24.67 &  $21.6 \pm 1.0$ & 22.1& $\cdots$  \\
\hline
\end{tabular}
\end{minipage}
\end{table*}
In the following we list the $H$-band limits obtained from our observations. 
Since the main focus of our observations has been relative astrometry no standard stars have been observed and sky conditions were generally not photometric. Furthermore, individual runs from different days have been combined to detect faint sources.
Zeropoints (and their errors) obtained during the standard calibration plan by ESO were taken to derive a rough photometric calibration. The zeropoints are from the same date as the individual runs (see Table~\ref{obs}) and were averaged according to the combination of the runs.

The \stf code was applied for the photometric measurements resulting in relatively small - nominal - magnitude errors.
Due to the observing conditions (usually clear or thin cirrus) and the combinations of individual runs the error is larger. 
In Table~\ref{hlimtab} we report the $H$-band limits at the positions of the neutron stars as well as the determined minimal photometric errors.  In case of a sufficient number (at least 5) of 2MASS sources with very good quality flags in all bands we compared our \stf magnitudes with the 2MASS values for the respective sources. A correction to our magnitudes was then applied as well as its additional magnitude error (standard deviation of the found differences).
For fields without sufficient 2MASS source number, an assumed magnitude error of 0.1 mag ($1\sigma$) was considered additionally to the \stf and zeropoint error.
This reflects the estimated uncertainties due to the sky conditions and the airmass correction for the respective fields. The average atmospheric extinction in the $H$-band at Paranal is 0.06~mag \footnote[1]{www.eso.org/instruments/isaac/tools/imaging\_standards.html\#BB}.
We list also the {\textsl{Source Extractor}} magnitudes derived by only doing relative photometry with the available, preferably faint, 2MASS sources. 
We regard these magnitudes as much coarser since often only very few 2MASS sources are present (e.g. only two sources for the field of PSR J0108-1431) and not all have good quality flags. Therefore we assigned a much higher magnitude error towards these "SE 2MASS-limits".
For some of our observations \citet{Curto2007} have already estimated $H$-band
limits (by relative photometry with all 2MASS sources in the field) which are listed for comparison in Table~\ref{hlimtab}.
 Except for  RX J0720.4-3125, where we have a slightly deeper limit due to
  the second epoch observation, our values correspond within the error bars to the results for 4 XTINSs investigated by
  \citet{Curto2007}.

If two epoch observations are available, it would be interesting if the combined image gives a deeper limit. 
In case of PSR J1932+1059 the $NACO$ observation improves the constraints in the $H$-band limit since the bright source close to the pulsar was behind the coronograph. 
In the $NACO$ field sources are detected down to $21.3$~mag. Since photons from the bright star are present close to the position of the NS, this might hinder detection of very faint objects. Therefore we take as limit the $H$ magnitude of the faintest source which has a similar angular separation from the bright source as the pulsar. Thus, the $H$-band limit for PSR J1932+1059 is $20.31 \pm 0.18$~mag.

\begin{figure*}[t]
\begin{minipage}[l]{.22\textwidth}
 \includegraphics[width=4.5cm,keepaspectratio=1]{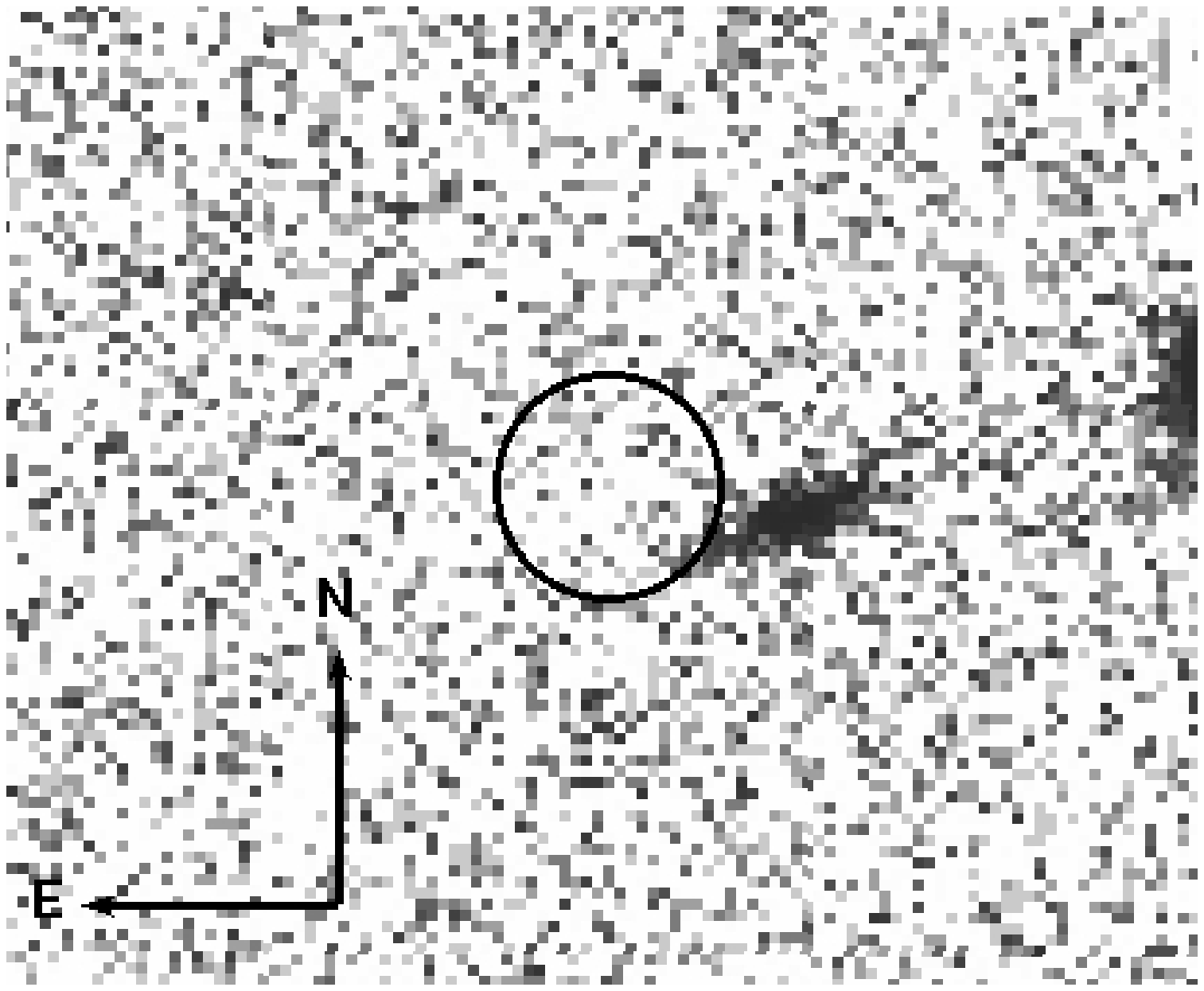}
\end{minipage}
 \hspace{0.5cm}
\begin{minipage}[l]{.22\textwidth}
 \hspace{-0.1cm}
 \includegraphics[width=4.5cm,keepaspectratio=1]{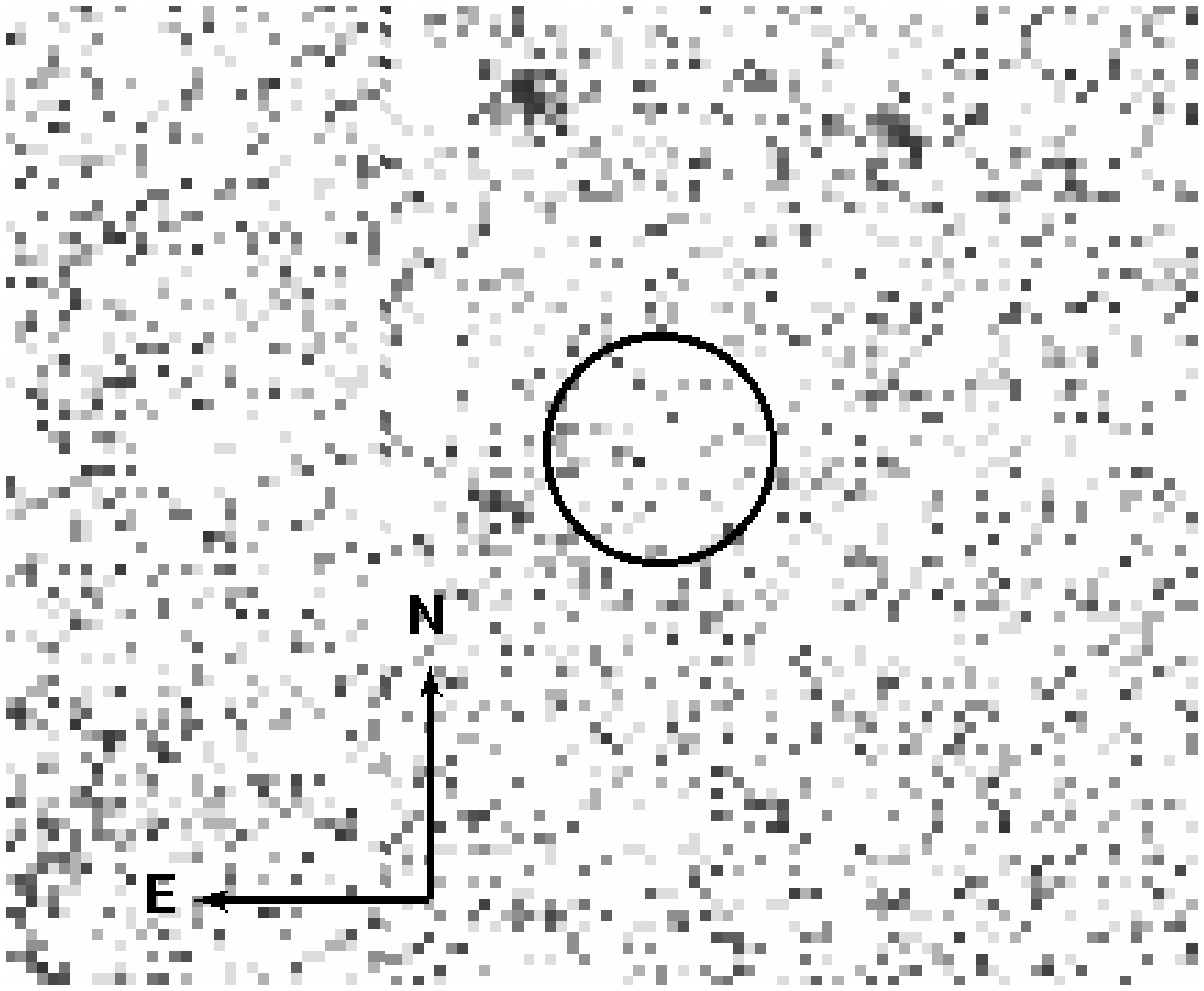}
\end{minipage}
 \hspace{0.5cm}
\begin{minipage}[l]{.22\textwidth}
 \includegraphics[width=4.5cm,keepaspectratio=1]{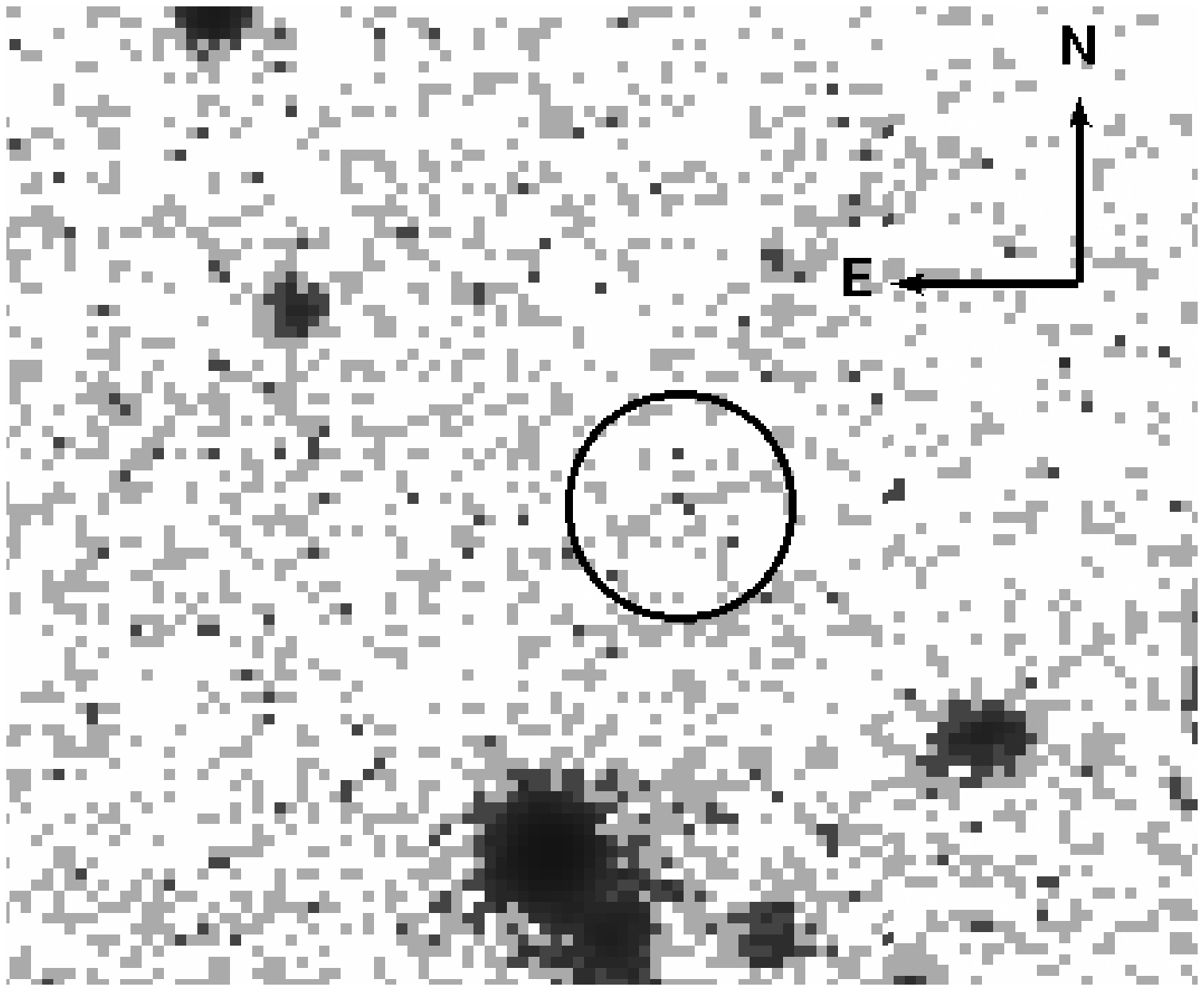}
\end{minipage}
 \hspace{0.5cm}
\begin{minipage}[l]{.22\textwidth}
 \includegraphics[width=4.5cm,keepaspectratio=1]{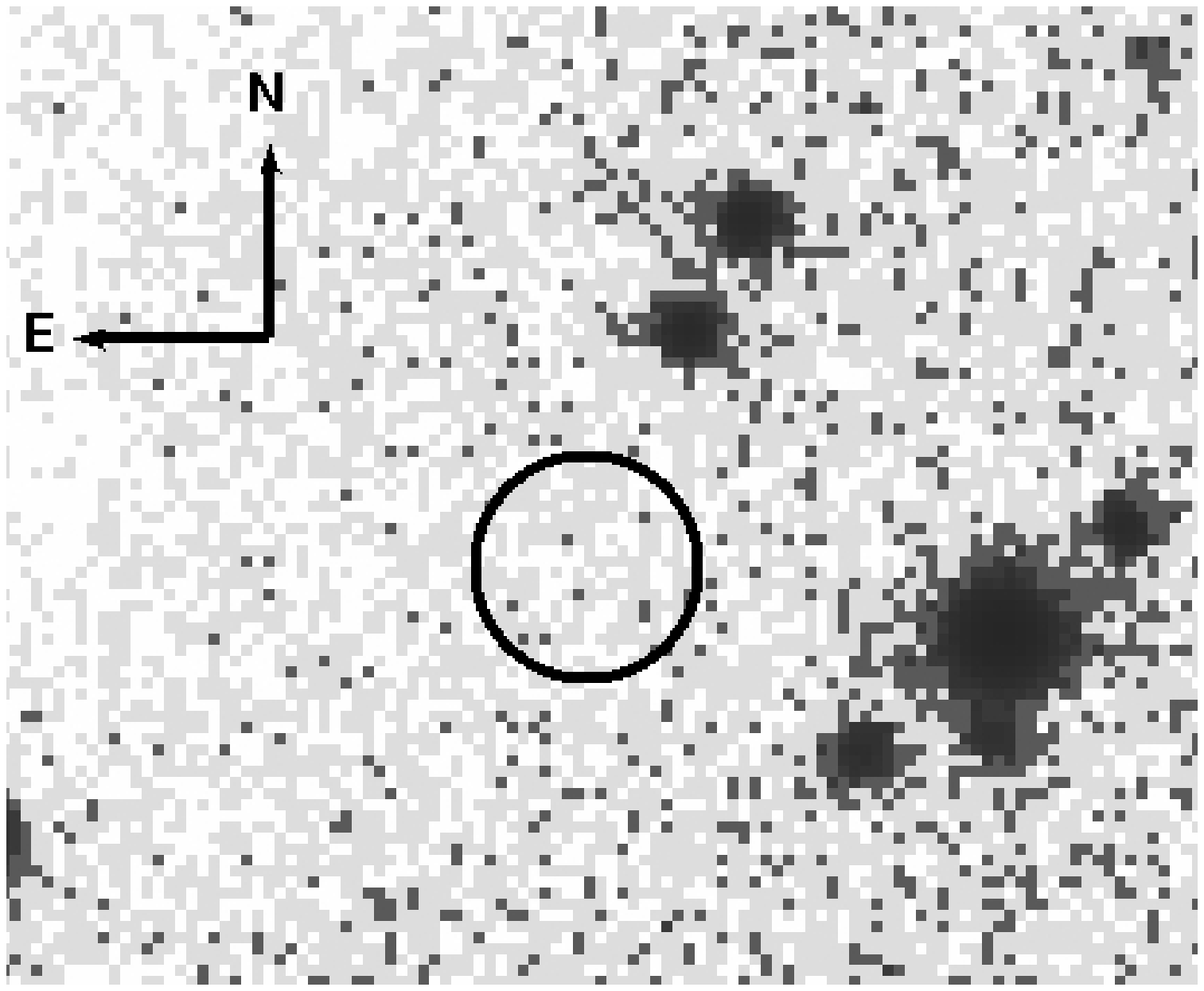}
\end{minipage}
\protect\\

\begin{minipage}[l]{.22\textwidth}
 \includegraphics[width=4.5cm,keepaspectratio=1]{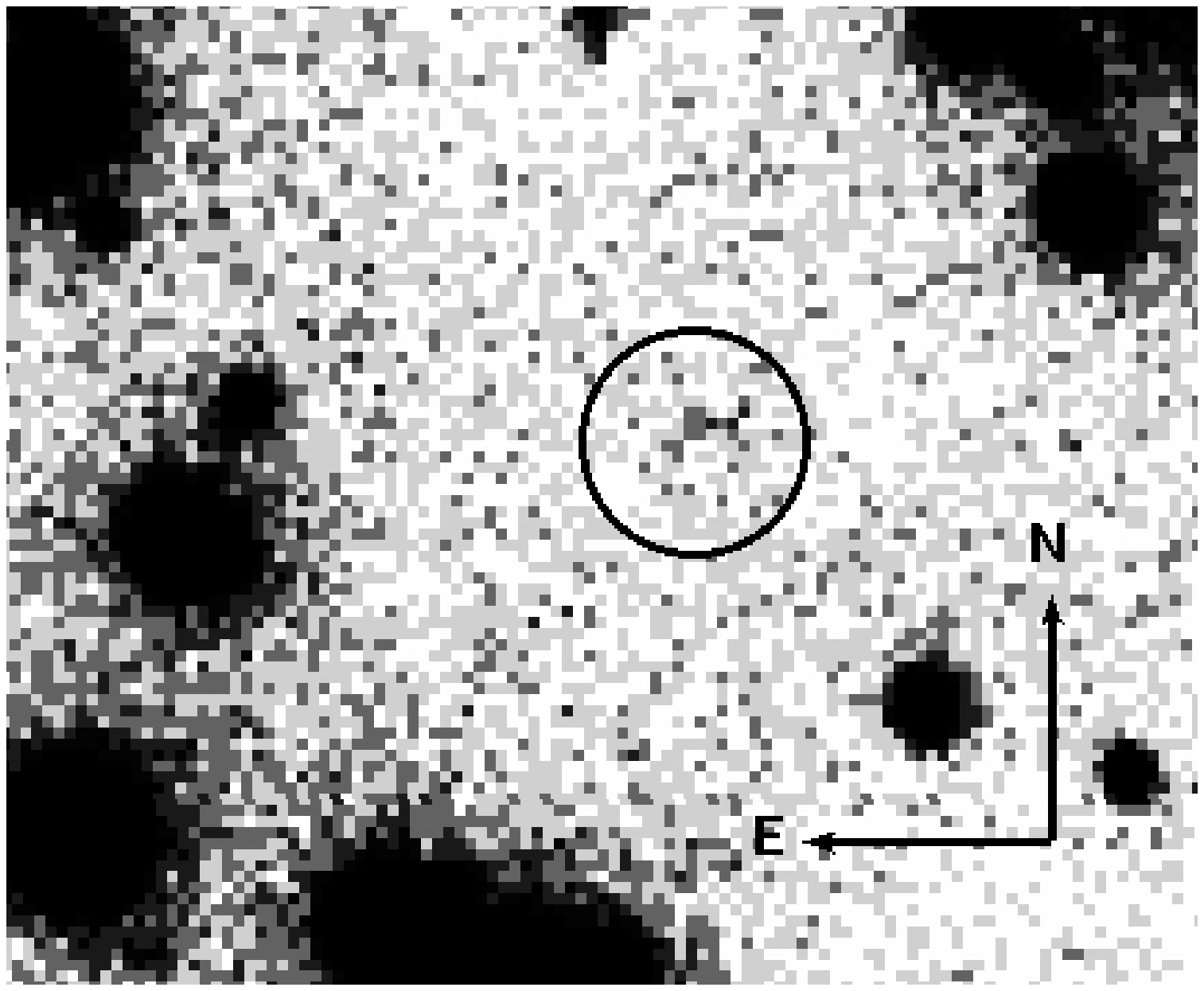}
\end{minipage}
 \hspace{0.5cm}
\begin{minipage}[l]{.22\textwidth}
 \includegraphics[width=4.5cm,keepaspectratio=1]{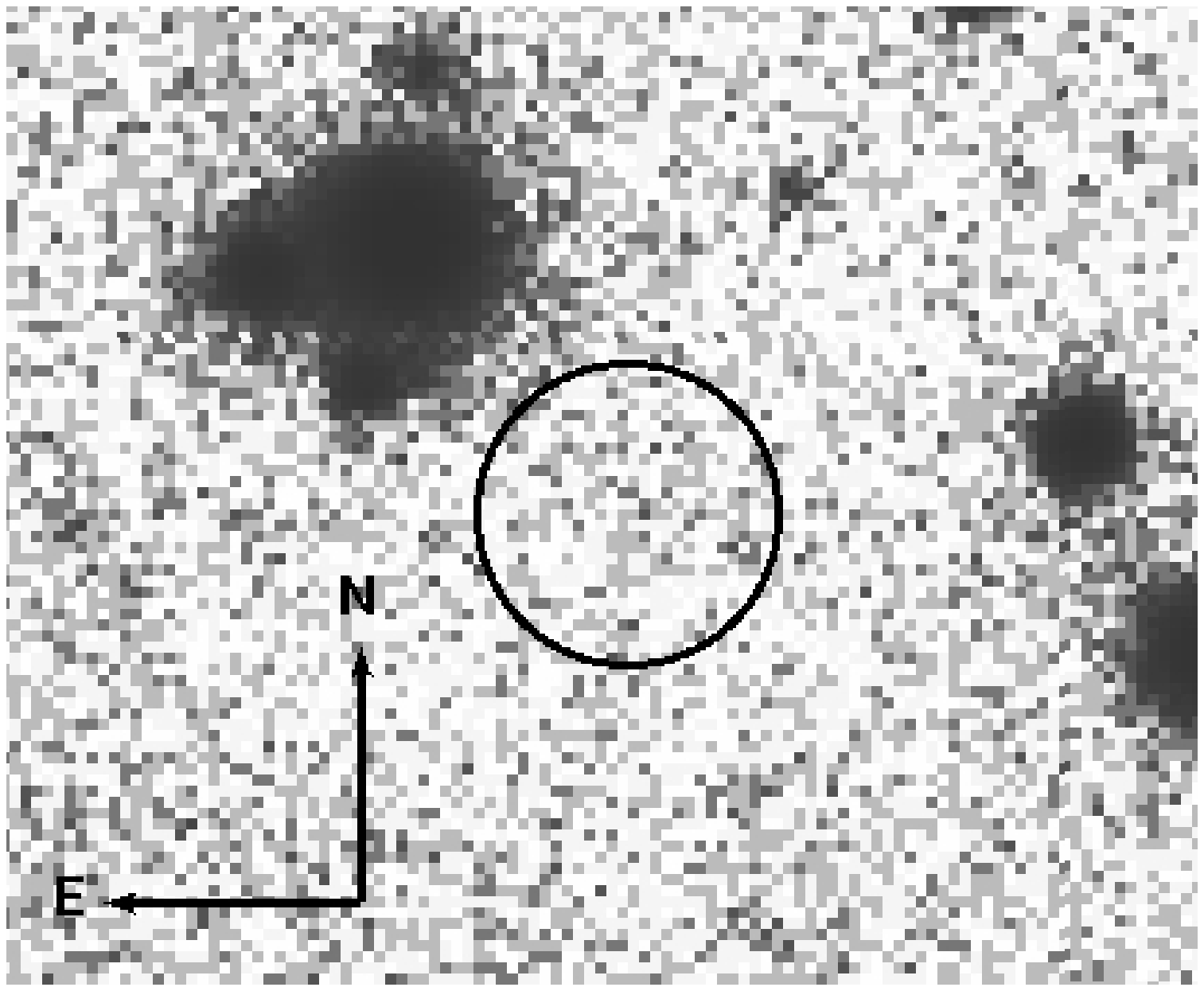}
\end{minipage}
 \hspace{0.5cm}
\begin{minipage}[l]{.22\textwidth}
 \includegraphics[width=4.5cm,keepaspectratio=1]{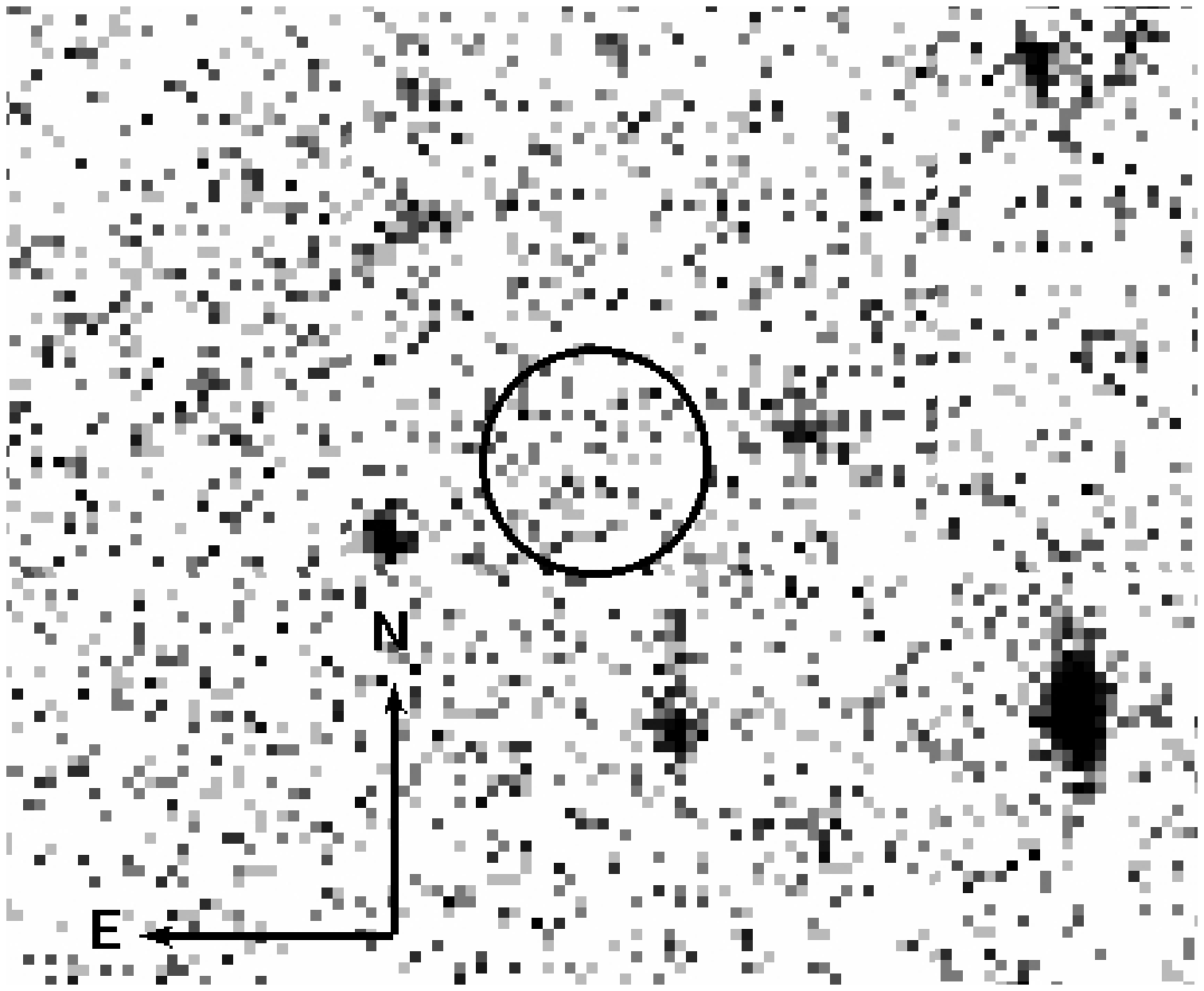}
\end{minipage}
\hfill
 \begin{flushleft}
 \caption{\textbf{Neutron star detection limits in the $H$-band} \protect\\}
Moving left to right, top to bottom the $H$-band fields of the following neutron stars are shown:
first row: PSR J0108-1431, RX J0420.0-5022, Geminga, RX J0720.4-3125;
second row: RX J0806.4-4123, RX J1856.6-3754, RBS 1774.
Except one each circle has a radius of 1.5\arcsec. The radius for the circle around the position of RX J1856.6-3754 has a radius of 2\arcsec.
The box sizes are roughly $16\arcsec \times 13\arcsec$. See also text. \vspace{-0.5cm}
 \label{fig:limits}  
 \end{flushleft}
\end{figure*}

In case of two $ISAAC$ epochs (similar observing conditions) we combined both to test for a deeper limit for Geminga and RX J0720.4-3125. The Zeropoints are averaged for the two epochs and \stf measurements were refined with high-quality-2MASS point sources as described above.
In case of RX J1856.6-3754 the $SOFI$ image was transformed to the $ISAAC$ reference frame and both images averaged.
The \stf magnitudes were first calibrated roughly with the $ISAAC$ zeropoint and then refined with 2MASS as well.
In Table~\ref{hlimtab}, we list the results of the combined observations.
In Figure~\ref{fig:limits}, the images around the NS's position are shown, in case of second epochs the combined image is chosen. The field of PSR J1932+1059 is shown in Fig.~\ref{field1932}.
For RX J0806.4-4123 one might suspect a very faint source within the $r=1.5\arcsec$ circle.
However, this is not an $3\sigma$ detection  ($3\sigma$ $H$-band limit is $22.64 \pm 0.16$~mag).
Deeper observations would be of interest here although the field of RX J0806.4-4123 is very crowded, and the chance of background objects is high.

\section{Discussion}

We analysed multiple-epoch $H$-band observations of four isolated neutron stars and found no co-moving object in our data. The derived limits in the frame of the B97 calculations are 12, 15, 11, 42 Jupiter masses for Geminga, RX~J0720.4-3125, RX~J1856.6-3754, and PSR~J1932+1059 assuming ages of 1, 1, 1, 3.1~Myrs and distances of 250, 361, 167, 361\,pc, respectively.
The small sample of investigated objects allows only crude statistical conclusions.
A new query at the most recent pulsar data base (\citealt{Manchester2005}
\footnote[2]{http://www.atnf.csiro.au/research/pulsar/psrcat}) for pulsars
with ages below 100~Myrs and distances below 400~pc, results in ten NSs
(including Geminga, and the XTINSs RX~J0720.4-3125 and RX~J1856.6-3754).
Listed among them is the double-pulsar PSR~J0737-3039. However, we exclude this system since first, the
distance stated in the corresponding reference \citep{Kramer2006} ranges from
200 to 1000~pc, and second, this would imply a merging of statistics for
double NS systems with isolated NSs.
We consider two XTINSs additional to RX~J0720.4-3125 and RX~J1856.6-3754 as
the upper limit for the number of XTINSs within 400~pc. This upper limit is
chosen because first, already the second brightest XTINS, RX~J0720.4-3125, has
a distance of 361~pc,  and second, we take into account the indications from
the absorbing hydrogen column density and ISM models for the distances of the
fainter XTINSs (e.g. \citealt{Posselt2007}),
Thus, we consider 11 NSs within 400~pc as young ($<100$~Myrs) enough to
show co-moving companions with our technique. We note that we do not include
the recently found Rotational Radio Transients, since the currently-identified
sources are thought to be at distances between 2 to 5~kpc \citep{Laugh2006}. 
 The probability, $P$, to get no detections in a subsample out of 11 NSs having an
(unkown) likelihood, $L_{Com}$, for  substellar companions, is equivalent to
an hypergeometric distribution probability. $P$ can be calculated according to
the conventional equations of the hypergeometric distribution for different
likelihoods $L_{Com}$.
Given 11 sources, a zero detection rate of brown dwarfs among four NSs results in
$P=$\,2\,\%, 21\,\%, or 64\,\%, if $L_{Com}$ is 55\,\%,  27\,\%, or 9\,\% , respectively.
For planets heavier than 15 Jupiter masses, the zero detection rate of
substellar companions among three NSs results in $P=$ 6\,\%, 34\,\%, or
73\,\%, if $L_{Com}$ is 55\,\%,  27\,\%, or 9\,\% , respectively.
It is likely that the young population of isolated NSs has a similar
substellar companion fraction as older isolated NSs.  
Strictly spoken, the numbers presented above cover only the case of
formation of substellar companions {\emph{after}} the supernova since we
  approximated the oldest age of possible substellar
  companions with the NS age. However, the SN survivor's rate is likely to be even smaller.
We conclude that brown dwarfs and giant gas planets heavier than 15 Jupiter masses are rare among isolated NSs.
Possible reasons are: 1. These substellar companions did not survive the destructive past of the NS. 2. They are too massive to be formed from a supernova fallback disk. 3. Massive giant planets ($>10$~M$_{Jupiter}$) are rare, at least among the well-studied low-mass stars. Less than 5\% of the known planets are such objects (March 2008, Schneider\footnote[1]{The Extrasolar Planets Encyclopaedia: http://exoplanet.eu/}).
4. Only few massive stellar progenitors host giant planets. There are indications that the rate of substellar companions decreases with stellar mass. For example, considering low-mass ($< 5$~M$_{\odot}$) stars, \citet{Kennedy2008} found a peak at stellar host masses $\sim 3$~M$_{\odot}$ for the probability of having at least one gas giant around them. While from massive star formation sites one knows disks around protostars with masses less than $\sim 20$~M$_{\odot}$ \citep{Cesaroni2007}, these disk are expected to have a shorter lifetime than those around low-mass stars and this may reduce the formation rate of massive substellar companions.
There is currently no planet known around a star more massive then
$4.5$~M$_{\odot}$ (March 2008, Schneider), which may be, however, partly due
to instrumental limitations (e.g. brightness contrast).
 5. Substellar companions may be fainter than calculated in B97.
Comparing to other theoretical models based on non-gray atmospheres, the B97 effective temperatures are hotter  \citep{Baraffe2002}. Thus, actual $H$-magnitudes might be lower than calculated by B97.
\\
Seeing our result in the light of the searches for substellar companions around white dwarfs it is interesting that substellar companions seem to be sparse among white dwarfs too:
 until now only three L-type companions among well over 1000 white dwarf targets are known (\citealt{Farihi2008} and references therein) and one suspected planet ($\sim 2$~M$_{Jupiter}$) around a pulsating white dwarf \citep{Mullally2008} has yet to be confirmed.\\

A short note on RX~J0720.4-3125: \citet{Haberl2006aph} suspected free
precession to be responsible for the $\sim 7$~years variations in the pulsed
X-ray spectra. Note, that this is still a subject under discussion \citep{Kerkwijk2007noprecision}.
In the frame of our work it is interesting to note, why forced precession, e.g., by a substellar companion, is not possible. 
For example, \citet{Liu2007} discussed forced precession by a planet for PSR B1828-11. In the case of RX~J0720.4-3125, we can apply the same formula as   
\citet{Liu2007} 
to approximate the orbital radius which would be needed for planets lighter than 15 Jupiter masses so that they could cause the precession observed. The largest orbital radius, such planets could have, would be $1 \times 10^{-5}$~AU (for zero inclination, NS distortion of $4 \times 10^{-8}$). However, this would be below the smallest Roche limit for a (normal) NS (radius of 10\,km, density of $10^{16}$\,kg\,m$^{-3}$) - both for planets with densities similar to Earth, resulting in a Roche limit of $2 \times 10^{-3}$~AU; and for planets with densities similar to Jupiter, resulting in a Roche limit of $3 \times 10^{-3}$~AU. 
Thus, even if there may be planets lighter than 15 Jupiter masses around RX~J0720.4-3125, they cannot cause a $\sim 7$~years precession of the NS. \\
 
In agreement with expectations no neutron star was detected in the $H$-band. 
The NIR limits are only weak constraints on the overall NS spectra given the known 
deep optical observations.  This result is in agreement
  with that of the independent investigation for the XTINSs by \citet{Curto2007}.

\begin{acknowledgements}
We thank Andreas Seifahrt for extensive discussions about source detection algorithm accuracy, Lowell Tacconi-Garman for suggestions regarding the $NACO$-observations, as well as the service astronomers at ESO for observing our sources.
Furthermore, we thank also P. Draper and E. Diolaiti for answering questions
regarding $GAIA$ and \stf.\\
We thank the anonymous referee for careful reading and constructive comments which helped 
improving the quality of this article. \\
This research has made use of SAOImage DS9, developed 
by SAO; the SIMBAD and VizieR databases,
operated at CDS, Strasbourg, France; and SAO/NASA's Astrophysics Data System Bibliographic Services.

\end{acknowledgements}

\bibliographystyle{aa} 
\bibliography{planetsBIB}

\Online

\begin{appendix} 
\section{Observations}
\label{detailsobs}

\begin{table*}
\caption[Observations]{\textbf{List of our $H$-band observations} \protect\\  
The night parameters of each observation are averaged. Note that the DIMM seeing is the standard ESO $V$-band value. The Near Infrared seeing is usually better by at least 0.1 arcsec.
In the column "Included ?" it is stated whether we took this observation run
into account or did not use it, e.g., because of known technical problems or
bad weather during the observations. Furthermore, it is noted by "Y-mod" if
individual files were excluded from the observational run, e.g. due to
exceptional bad seeing.
For completeness we note that five neutron star fields were already observed in December 2000 using
 ESO's NTT equipped with the $SOFI$ instrument in the  $H$-band (66.D-0135). However, comparing to the
later VLT $ISAAC$ observations, the $SOFI$ images were generally
less deep and suffered from worse seeing conditions. 
We do not list the NTT $SOFI$ observations from 2000
here.}
\label{obs}
\begin {tabular}{@{} l   c  c  c  c  c  c  c @{}}
\hline
Object & Instrument & Exposure & Nights & Airmass & DIMM Seeing &  Sky &  Included ? \\  
       &            & [s]     & yyyy-mm-dd &    & [arcsec] &    &  Yes / No  \\ 
\hline
PSR J0108-1431 & VLT $ISAAC$ & 1200 & 2003-06-23 & 1.1 & 0.7 & photometric & Y-mod \\ 
               & VLT $ISAAC$ & 1380 & 2003-08-07 & 1.0 & 0.8 & photometric & Y-mod \\ 

RX J0420.0-5022 &VLT $ISAAC$ & 1980 & 2004-01-10 & 1.3 & 0.6 & clouds & N \\ 
                &VLT $ISAAC$ & 1980 & 2004-01-13 & 1.7 & 0.9 & photometric & Y-mod\\ 
                &VLT $ISAAC$ & 1980 & 2003-11-13 & 1.4 & 0.8 & clear & Y\\

Geminga   &VLT $ISAAC$ & 1980 & 2004-01-02 & 1.6 & 0.9 & photometric & Y \\           
          &VLT $ISAAC$ &  900 & 2003-12-30 & 1.4 & 0.6 & cirrus & Y \\           
          &VLT $ISAAC$ & 1980 & 2003-11-15 & 1.4 & 0.5 & clear & Y  \\                               

RX J0720.4-3125 &VLT $ISAAC$ & 1980 & 2004-01-08 & 1.0 & 0.5 & cirrus & Y \\
                &VLT $ISAAC$ & 1980 & 2004-01-11 & 1.0 & 0.7 & clear & Y \\
                &VLT $ISAAC$ & 1980 & 2003-12-04 & 1.0 & 0.8 & photometric & Y \\

RX J0806.4-4123 &VLT $ISAAC$ & 1980 & 2004-12-21 & 1.1 & 0.7 & photometric & Y \\
                &VLT $ISAAC$ & 1980 & 2004-12-21 & 1.1 & 0.6 & photometric & Y \\
                &VLT $ISAAC$ & 1980 & 2004-11-25 & 1.1 & 0.7 & clear	& Y \\	

RX J1856.6-3754 & $ISAAC$ & 1380 & 2003-05-22 &1.1 & 0.7 & clear & Y \\
                & $ISAAC$ & 1380 & 2003-05-22 &1.1 & 0.7 & clear & Y \\

PSR J1932+1059  &VLT $ISAAC$ & 1380 & 2003-06-05 & 1.3 & 0.6 & clear & Y \\
                &VLT $ISAAC$ & 1380 & 2003-06-04 & 1.3 & 0.8 & clear & Y-mod\\
                &VLT $ISAAC$ & 1380 & 2003-06-17 & 1.7 & 0.7 & clear & Y \\
		
PSR J2124-3358  &VLT $ISAAC$ & 1380 & 2003-06-05 & 1.0 & 0.5 & clear & Y \\
                &VLT $ISAAC$ & 1380 & 2003-05-22 & 1.0 & 0.7 & clear & Y-mod \\
		
RBS 1774        &VLT $ISAAC$ & 1200 & 2003-06-06 & 1.2 & 0.6 & clear & Y \\
                &VLT $ISAAC$ & 1380 & 2003-06-06 & 1.2 & 0.5 & clear & Y \\

\hline
 \multicolumn{8}{l}{Second Epoch Observations} \\
\hline
Geminga   &VLT $ISAAC$ & 2040 & 2007-02-08 & 1.5 & 0.8 &  clear   & Y-mod \\
          &VLT $ISAAC$ & 2040 & 2007-02-08 & 1.7 & 0.7 &  clear   & Y-mod \\
          &VLT $ISAAC$ & 2040 & 2007-01-26 & 1.7 & 0.5 & photometric& Y \\
          &VLT $ISAAC$ & 2040 & 2007-01-27 & 1.4 & 0.7 & photometric& Y \\

RX J0720.4-3125 &VLT $ISAAC$ & 1980 & 2006-12-27 & 1.3 &0.8 & clear   & Y \\
                &VLT $ISAAC$ & 3420 & 2006-12-27 & 1.1 &0.9 & clear   & Y-mod \\
                &VLT $ISAAC$ & 2040 & 2006-12-25 & 1.1 &0.8 & clear   & Y \\
                &VLT $ISAAC$ & 2040 & 2006-12-25 & 1.1 &0.9 & clear   & Y \\

RX J1856.6-3754 &NTT $SOFI$ & 24240 & 2006-05-09 & 1.3 & 0.8 & clear & Y-mod  \\
                &NTT $SOFI$ & 15960 & 2006-05-10 & 1.4 & 0.7 & cirrus & Y-mod  \\
                &NTT $SOFI$ & 9120 & 2006-05-11 & 1.5 & 0.8 & cloudy & Y-mod  \\

PSR J1932+1059  &NTT $SOFI$ & 7200 & 2006-05-10 & 1.4 & 0.7 & cirrus & Y-mod  \\
                &NTT $SOFI$ & 14400 & 2006-05-11 & 1.6 & 0.8 & cloudy & Y-mod  \\
		&VLT $NACO$ & 1360 & 2007-05-02 & 1.3 & 0.6 & clear & Y-mod \\
                
\hline
\end{tabular}
\end{table*}

The VLT {\it{NAos COnica}} ($NACO$) coronographic observations of
PSR~J1932+1059  were done with the S54 camera  and the opaque $1.4\arcsec$ coronograph.
As a reference star for the AO correction the bright ($V=13.4$) star GSC2
N0232103780 was selected, located $~ 5\arcsec$ away from PSR~J1932+1059.
A jitter box of $20\arcsec$, coronographic individual exposures of 60\,s, sky individual exposures of 0.3454\,s, 4 "object" offset positions and 6 "sky" offset positions  were further parameters of this $NACO$ run.
The $jitter$ engine of $Eclipse$ was applied for data reduction
considering the corresponding  dark and lamp flat, 28 individually
selected objects for the image stacking algorithm and separate object and sky
files.
 
Due to the relatively small number of images, the leaking-in photons of the
bright object outside the actual field of view, and the coronograph wire
construction, the resulting image (Fig.~\ref{field1932}) shows several faint holes and artefacts which may result in false
source detections. We inspected each found source
and excluded such artefacts from further analysis. Since we are not interested
in highly accurate photometry but in (relative) astrometry the image quality
serves its intended purpose.\\  

\section{ Further details on methods}
\label{detailsm}
\subsection{Other source detection algorithms}
For the relative astrometry, we considered two different approaches.
First, we applied $GAIA$ \citep{gaia} for object detection provided that a signal-to-noise ratio (SNR) of at least three is reached. $GAIA$ makes use of the {\textsl{Source Extractor}} by \citet{Bertin1996}. In this program, which was designed with barycenter-detection of galaxies in mind, one can choose several detection parameters like the used filters (e.g. Gaussian), or how to handle overlapping sources. We tested several sets of parameters and applied individually the apparently best suited one for each observational field.  
We then estimated the $H$-band magnitude by calibrating with the {\textsl{Source Extractor}} (preferably faint) 2MASS PSC sources.
The achieved SNR for a 22~mag object was around three, magnitude errors were around 0.4 mag. 
The positional accuracy for faint sources stated by the {\textsl{Source Extractor}} was often surprisingly good (better than one tenth of pixel for faint sources) which seems not realistic. 
We checked the positional accuracy reached by 2D-Gaussfits using $IDL$. 
For the interesting -- usually faint -- objects, three times the mean positional uncertainty was $3\sigma \approx 80$ mas, but could also reach exceptionally $230$ mas for the faintest. The pixel scale of $ISAAC$ was $\approx 147$ mas.
Brighter objects had usually better determined positions. While the 2D-Gaussfits were useful to illustrate the possible maximum positional errors, they failed for overlapping sources.
Therefore, we applied as final approach for the relative astrometry the
\stf-code by \citet{Diolaiti2000SPIE} with its known accuracy as noted in Sec.~\ref{method}. 
For bright $H$-band objects the $3 \sigma$ error of the pixel shift is around
one hundredth of a pixel, while for the faintest object it reaches one pixel.
We compared the {\textsl{Source Extractor / GAIA }} source list with the \stf
source list for a performance check. \stf was usually able to find at least as
many or more sources as the {\textsl{Source Extractor}} of $GAIA$, and it only
failed for galaxies which are excluded anyway from further analysis.

\subsection{Procedure in case of first epoch only observations }
\label{detailsno2}
If no second epoch observation was obtained, e.g., due to still unknown proper
motion, we used the following procedure to identify at least potentially
interesting objects for follow-up observations. Orbital separations of at
least 5000~AU are inspected for NIR sources fainter than the magnitude
obtained by the luminosity-magnitude relation and considering B97 at the brown
dwarf upper mass limit. NS distances listed in Tab.~\ref{objects} were used to
determine the 5000 AU circle around the NS position.  The relatively wide
separations take into account recently found distant substellar companions at
roughly  4000~AU around main sequence stars (see e.g. table 3 in
\citet{Burgasser2005} for brown dwarf companions) and possible widening of orbits
due to the mass loss of the primary during the supernova according to Kepler's laws.

The faint sources detected by $GAIA$ were cross-correlated to sources in available optical observations obtaining optical-NIR-colors. 
This was usually done with the help of archival or kindly provided optical
observations. Many first-epoch-objects could be rejected as possible
substellar companion candidates due to these colors being typical for normal
stars according to \citet{Kenyon1995}. The given optical and
optical-NIR-colors by \citet{Kenyon1995} go down to spectral type M6. As an
example, one would expect the $R-H$ color to be lower than 5~mag for normal
stars down to M6~V, while L and T dwarfs usually have higher values of around
6~mag. This can be seen in Fig.~1 of \cite{Kirk1999} shown for the $R-K$ color
(the $H-K$ colors are around 0.6~mag for most L and T dwarfs, see e.g. Fig.~14
of \citealt{Kirk1999}). However, we could not exclude ultra-cool M-dwarfs (M8)
at this stage.

We note the use of a conventional magnitude error of usually 1~mag up to
exceptionally 2~mag for the optical-NIR-colors considering variations in the
magnitudes of the reference optical observations as well as errors in the
magnitude estimation of the NIR-objects. By this, all potentially interesting
sources should be included. \\

\newpage
\section{PSR J2124-3358 }
\label{psr2124}
\begin{figure}[b]
 \psfig{figure=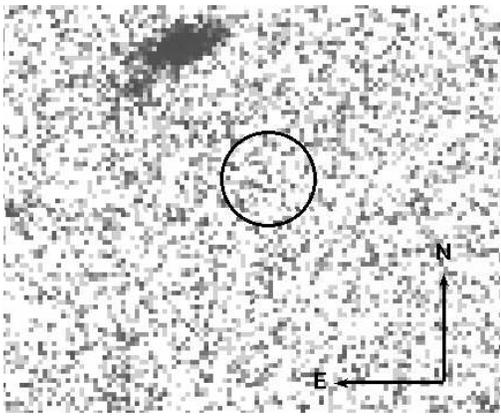,width=6.6cm}
  \caption{{\textbf{Detection limit in the $H$-band for PSR J2124-3358}}
The radius for the circle around the position of the NS has a radius of
1.5\arcsec. The box size is  roughly $16\arcsec \times 13\arcsec$. }
  \label{fig:2124limits}  
\end{figure}

PSR J2124-3358 is a millisecond radio pulsar. The up-to-date ATNF pulsar
catalogue \citep{Manchester2005} indicates an age of $5.86 \cdot 10^9$ years
which is considerably older than we thought when selecting this source some
years ago from \citet{Taylor1993} in its latest online version (no age for PSR
J2124-3358 noted). At this age and a distance of 270~pc, cooling substellar
companions would have magnitudes fainter than $H=24.9$~mag following
\citet{Burrows1997}.
Our $ISAAC$ $H$-band observations are not deep enough to detect such objects.
We derived the NIR limit at the position of the NS which is $H_{lim} \pm 1
\sigma = 21.22 \pm 0.16$\,mag, the area around the NS is shown in Fig.~\ref{fig:2124limits}.

\end{appendix}
\end{document}